# Topological Pseudospin Hall Effect and Multi-frequency Corner Modes in Kagome-based Lattices

*Shenglong Guo[1*], Qinhui Jiang[1*], Yuma Kawaguchi[2], Bo Li[1,3], and Mengyao Li[1†]*

[1]*Tsinghua Shenzhen International Graduate School, Tsinghua University, Shenzhen 518055, China*

[2]*Department of Electrical Engineering, The City College of New York, New York, NY 10031, USA*

[3]*Suzhou Laboratory, Suzhou 215000, China*

*\*These authors contributed equally.*

*[†]Contact information: mengyaoli@sz.tsinghua.edu.cn*

**ABSTRACT:**

Topological phases and modes, including pseudospin-Hall–selective edge transport and corner states, provide robust control of wave propagation and modal confinement in classical wave platforms. Under a tight-binding framework, we theoretically investigate two lattice designs derived from the kagome lattice. These extended kagome lattices support a series of localized modes, including pseudospin-Hall-like topological edge states and corner modes in different bandgaps and frequencies, which were not only achieved under lower lattice symmetries than Wu-Hu lattices, but also enable more degrees of freedom in topological and localized modes. By introducing two types of extended kagome lattices with different topological properties, multiple interesting phenomena, including newly emerged multiple groups of corner states, parametric tunable pseudospin Hall effect, and type-II corner states without long-range interactions, are found in theoretical models, which are possible and viable to achieve in artificial classical systems such as photonic, acoustic, or electrical circuits.

## I. INTRODUCTION.

Topological phases give rise to a variety of new phenomena and novel applications in not only condensed matter systems, but also in a wider range of platforms, including photonics[1-14], acoustics[15-18], mechanics[19,20], and electronics[21-25]. In classical systems, leveraging time-reversal symmetry and avoiding extreme requirements such as strong magnetic bias, non-magnetic topological phases, including pseudospin-Hall[21,26-29] and valley-Hall [26,30]phases, have attracted broad interest. These phases enable direction-selective and robust guiding by pseudospin–orbit coupling, or by valley contrast when K-point momentum is preserved. [31]. In pseudospin Hall systems, the analogy between two-dimensional Dirac physics in quantum materials and classical platforms guides the control of Dirac bands and edge states, enabling robust wave guiding[32], pseudospin engineering[5,33-37], helical coupling and lasing[38,39].

Unlike the standard bulk boundary correspondence, which induces topological boundary modes one dimension lower than bulk, higher order topological insulators (HOTIs) host protected modes on boundaries of lower dimensions, for example, hinge states in three-dimensional systems[40,41] and corner states in two or three dimensions[42,43], and more interestingly, leading to high quality factor corner localized modes. Among the topological corner modes, type-I corner states are modes localized at geometric corners, which are reported in many higher-order topological insulators[44-46], and type-II corner states arise from edge bands and typically require couplings of next nearest neighbor and beyond to form and to be observed[47-49]. However, whether there exist alternative approaches without long-range interactions or local perturbations to achieve localized modes that distribute like type-II corner states is interesting and challenging. Another interest of corner modes lies in multi-frequency and multigap modes in one system, while recent research demonstrated its possibility[50], the number of modes is still quite limited.

Motivated by these contexts, there is a strong interest in robust topological platforms that admit multiple localized modes together with design space supporting richer phenomena, yet realizing such systems remains challenging. The kagome lattice attracted our attention as a strong platform with rich phenomena induced by its band structures, and has been widely employed to realize topological valley Hall physics. We theoretically propose two new geometries of creating pseudospin-selection topological waveguiding based on the platform of kagome lattice, which constructs 2-fold degeneracy of Dirac cones at the $\Gamma$ point, opening the opportunity of rich topological physics, such as adiabatic waveguiding metasurfaces[51] and tunable guided modes. Moreover, these platforms of extended kagome systems support rich topological phases and modes, such as multiple groups of localized corner modes, in which type-II corner states emerged in one of the systems even without the existence of long-range couplings. The other system we investigated took advantage of parametric tuning and achieved accidental degeneracy of double Dirac cones, enabling topological phases in a system that originally did not support them. These platforms open possibilities for robust on-chip photonic circuits, enabling higher integration density and potentially multifunctional topological lasers[52,53] that exploit different edge or corner modes as distinct channels.

## II. RESULTS

### A. Topological Pseudospin Hall Effect in Extended Kagome lattices

Kagome lattice consists of 3 sites per unit cell, forming a triangular shape, and its band structure has a flat band and two Dirac cones at the K and K' valleys, respectively. Kagome lattice and breathing kagome lattice provide a rich platform of topological phenomena and applications, such as the flat band[54-56], the valley Hall effect and valley selective edge modes[57,58], being able to integrate with non-Hermitian topological properties[59,60], and higher-order topology[18,47,49,61,62], which give rise to a lot of attention in the topological regime.

Enlarging the real space unit cell by grouping multiple primitive cells and adding more sites reduces the reciprocal lattice vectors, causing the Brillouin zone to shrink and the bands to fold. Following this idea, we extend the kagome lattice to a larger unit cell and fold the K points to the $\Gamma$ point of the new unit cell in the reciprocal space. Here, we propose 2 types of geometries by applying different ways of grouping and extending the kagome lattice. Both geometries consist of 3 primitive unit cells of unperturbed kagome, resulting in 9 sites per unit cell, and we named them as ext-kagome-I and ext-kagome-II with distinct band structures and topological properties we introduce below. Both of these two methods induced kagome band structure folding, and their Dirac cones, which originally lay in K and K' valleys, now both moved to and meet at the $\Gamma$ point.

The configuration of the unit cell of the first type of extended kagome (ext-kagome-I) is shown in Fig. 1(a), by grouping 3 adjacent kagome unit cells in a triangular shape. A reference band structure of a manually folded unperturbed kagome lattice induced by unit cell enlargement is demonstrated in Fig. 1(b). Note that the folding of the Brillouin Zone here is constructed manually for comparison with perturbed band structures, and its properties, such as degeneracies and number of bands, do not reflect the physical properties of the unperturbed kagome lattice. Next, we introduce lattice perturbation to break translational symmetry and induce degeneracy breaking of the Dirac point. The strategy of lattice perturbation is increasing/decreasing all couplings of intracell couplings uniformly, which results in an opened band gap, while still keeping the 2-fold degeneracy of Dirac cones at the Γ point. We then describe the band structures of the ext-kagome-I lattice by tight tight-binding model (TBM). When perturbation of intracell/intercell $(K, J)$ couplings is imposed on the ext-kagome-I, the double Dirac band degeneracy is clearly present in Fig. 1(c) with lattice coupling parameters of $K = -2.0, J = -1.0$.

The double-degenerate Dirac cone at the Γ point serves as a fundamental prerequisite for realizing the pseudospin Hall effect, which is a classical system analog to the Kramers degeneracy in time-reversal-invariant fermionic systems. By observing the band structure of the unit cell for ext-kagome-I and its characteristic of Dirac cones being double-degenerated at the Γ point, we examine whether the lattice can form pseudospin Hall edge states. We first verified the topological properties of the structure through the effective Hamiltonian[63] to calculate $Z_2$ invariants[64] with similar methods used in the analytical formalism of Wu-Hu structure[65]. The detailed analysis process could be found in Section III of the manuscript. Here, the $Z_2$ invariant is defined in analogy to the Bernevig–Hughes–Zhang model, where a value of 1 (0) corresponds to the presence (absence) of helical edge states within the same lattice configuration. Based on the analysis, we can see that the phase of expanded structures ($J < K < 0$) is regarded as topological nontrivial, and the phase of shrunk structures $(K < J < 0)$ is regarded as trivial according to the $Z_2$ invariants. The equation determining the value of the $Z_2$ invariant is as follows.

$$C_1 = \frac{1}{2}(sgn(K - J) + 1) \quad (1)$$

Where $C_{1st} = 1$ represents the nontrivial phase and $C_{1st} = 0$ represents the trivial phase.

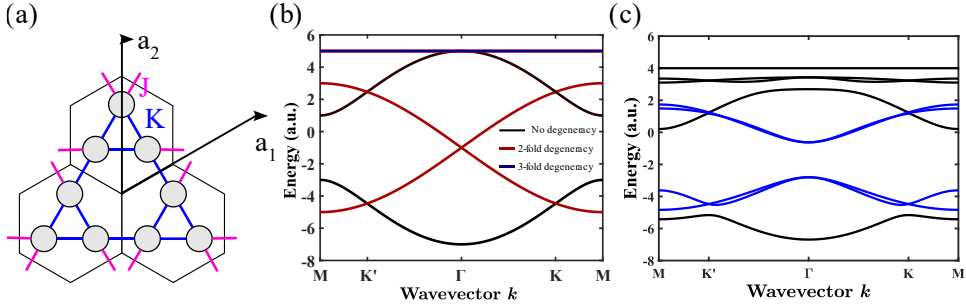

FIG. 1. Unit cells with different intracell coupling forms of ext-kagome-I. (a)The geometry of ext-kagome-I. Blue lines represent intracell couplings, and magenta lines represent intercell couplings. (b) Manually grouping 3 kagome unit cells and BZ folded of an unperturbed kagome lattice with 9 sites per unit cell, where the configuration is as (a) but for the J = K case. Color shows different degrees of degeneracy. (c) Dirac cone degeneracy breaking by introducing unbalanced intracell and intercell coupling. The intracell couplings have the same strength within each unit cell. Blue bands are highlighted as a pair of Dirac spin bands which degenerate at the Γ point $(K = -2.0, J = -1.0)$.

To visualize the edge band structures and mode distributions, we investigate the supercell in Fig. 2(a), which contains three ext-kagome-I lattice domains set to different intracell to intercell coupling ratios, forming a sandwich structure corresponding to the trivial-topological-trivial configurations. The middle layer of the supercell consists of topologically nontrivial ext-kagome-I structures $(K = -1.0, J = -1.5)$, upper and lower sides are trivial ext-kagome-I structures $(K = -1.5, J = -1.0)$. Each layer consists of 12 unit cells with the same parameters. The edge states (blue lines) associated with the pseudospin Hall effect are marked out in Fig. 2(b), which also have a small bandgap between edge state bands like the Wu-Hu structure[29]. The edge states are doubly degenerate due to the sandwich-like configuration, which forms two interfaces. The mode profile of the edge states, referring to the spatial distribution of the Bloch eigenvector amplitudes in the TBM, localized at the boundaries between topological non-trivial domains and trivial domains, is depicted in Fig. 2(c). The sandwich structure has two interfaces with topological phase transitions. The insets in Fig. 2(c) indicate the mode profile of the 2 unit cells at the boundaries between regions of extended kagome lattices with different topological phases.

    Due to the clear modification method of the lattice, the structure has the potential to be another adiabatic platform like the Wu-Hu lattice, which has been proven to hold better quality modes even in radiative regimes, and support a series of high-quality chiral bulk modes. The adiabatically linear transition of the topological interfaces between unit cells in a supercell is an emerging method that separates bulk modes to decrease the gap of the edge modes, thereby achieving an almost perfect quantum pseudospin Hall effect[51,66]. We construct another supercell in Fig. 2(d) that consists of unit cells adiabatically varying their intercell and intracell couplings. The unit cells from top to down are linearly transitioned gradually from expanded through balanced to shrunk. The band structure of the supercell in Fig. 2(e) has a superior effect with the smaller edge bandgap, whose size is approximately 3/10 of the edge bandgap shown in Fig. 2(b), further approaching the ideal quantum pseudospin Hall effect waveguide modes. Since only one interface is formed during adiabatic variation, the edge states in Fig. 2(e) are not doubly degenerate, which differs from those in Fig. 2(b). Meanwhile, the profile of the waveguide mode with adiabatically changing localized at the interfaces is demonstrated in Fig. 2(f). The insets in Fig. 2(f) show the interface of ext-kagome-I with expanded, balanced, and shrunk unit cells. Compared to the abrupt interface transformation of the step domain mode profile, the adiabatic linear interface transformation increases the width of the localized domain, which is also tunable with the gradient and profile of adiabatic parameter tunings. The possibility of constructing adiabatic metasurface platforms provides possibilities to design new spin waveguide devices with less radiative loss.

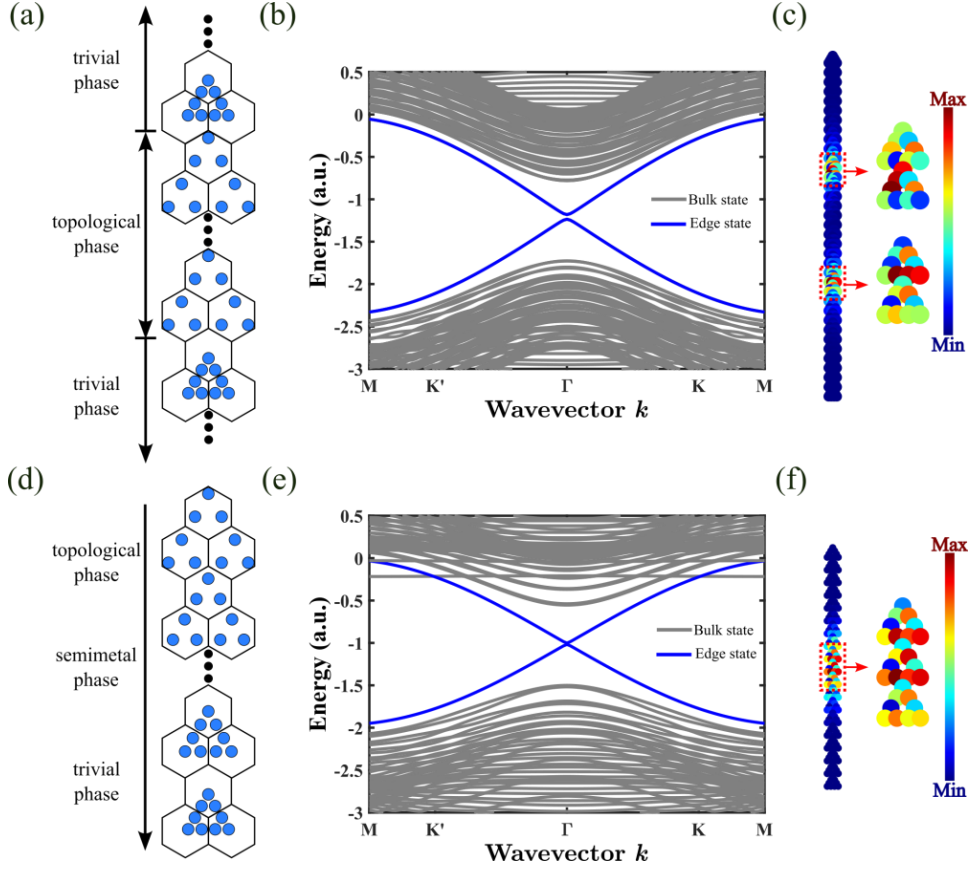

FIG. 2. Supercell of ext-kagome-I. (a) Geometry demonstration of supercell composed of two kinds of ext-kagome-I that consists of topological extended kagome lattice($K = -1.0, J = -1.5$) as the middle layer, upper and lower layers are trivial shrunk structures($K = -1.5, J = -1.0$). (b) Band structure of the supercell in (a). Blue lines highlight the edge bands, and gray lines represent the bulk bands. (c) Mode distribution of the edge states in (b), the inset image on the right is an enlarged zoom in interfaces between the topological and trivial domains. The amplitude of the mode profiles is color-coded in the figure. (d) Geometry demonstration of the supercell with adiabatically variable position of sites in unit cells. Unit cells from top to down are linearly varied from topological to trivial phase. (e) Band structure of the supercell in (d). Blue lines represent the edge bands and gray lines represent the bulk bands. (e) Mode profile for the edge state in (e), the image on the right is the enlarged interface between the topological phase, Dirac semimetal, and the trivial phase. Color bar showing the amplitude of the mode profiles.

    The second type of extended kagome unit cell (ext-kagome-II) comprises two upper and one lower conventional kagome unit cells, whose geometry is shown in Fig. 3. The most intuitive way of applying symmetry-breaking perturbation is the same method as we used for ext-kagome-I, where all sites shrink or expand uniformly. Under this modulation, all sites directly expand or shrink proportionally relative to the center of the unit cell [Fig. 3(a)], enabling the system to have all intracell coupling strengths the same. Similarly, we use the TBM to characterize the ext-kagome-II and calculate the band structure of the system. However, from the band structure of the ext-kagome-II in Fig. 3(b), we find that the pseudospin degeneracy at the Γ point is also lifted when such perturbations exist, which does not satisfy the conditions for the pseudospin Hall effect.

    Another strategy is tuning the coupling strengths of certain intracell couplings differently, while applying

these couplings with constraints such that the $C_{3v}$ symmetry is kept, as shown in Fig. 3(c). We discover that we can achieve accidental double Dirac cone degeneracy at the Γ point by certain combinations of values, which is the prerequisite for the pseudospin Hall effect. As a result, the band structure under certain combinations of parameters, in this case we chose $K_1 = -1.1$, $K_2 = K_3 = -0.6$, $K_4 = -1.8$, $J = -1.9$, can achieve double Dirac band degeneracy at the Γ point according to tight binding calculations [Fig. 3(d)]. Meanwhile, we find that the combinations of parameters that meet the condition of constructing double Dirac band degeneracy at the Γ point are not unique, such that multiple combinations could achieve this condition in the parameter space with distinct band structures, which opens the possibility of the construction of the topological and trivial domains (See Section II of Supplement Material[67]).

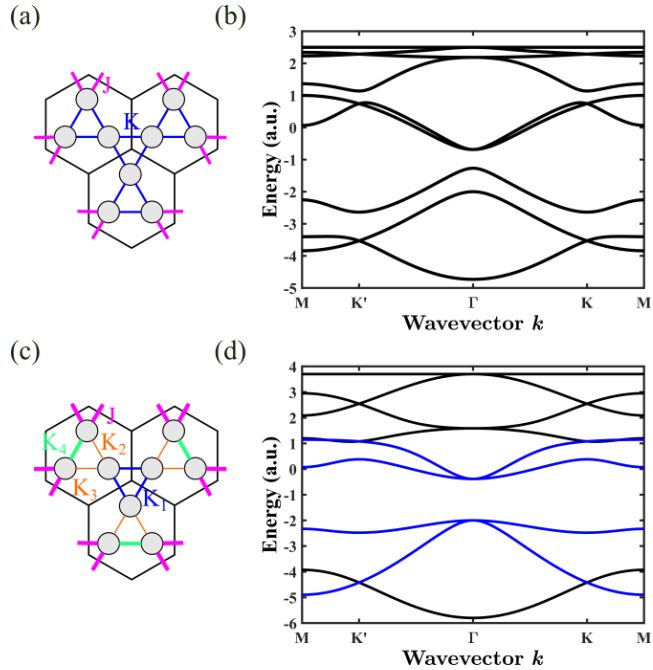

FIG. 3. Unit cells with different intracell coupling forms of ext-kagome-II. (a) The coupling of the ext-kagome-II lattice, whose intracell coupling strength is the same. (b) The energy band structure of the geometry in (a), in which the doubly-degenerated Drac cone disappears. (c) The coupling of the ext-kagome-II lattice, whose intracell coupling strength is tunable under certain constraints. $K_2 = K_3$ is one of the conditions that need to be satisfied. (d) The energy band structure of the geometry in (c). Blue bands are highlighted as a pair of Dirac spin bands that degenerate at the Γ point. Different line colors between nearest neighbor sites of (a) and (c) represent different kinds of coupling, and line thickness represents the strength of coupling.

We employ a similar sandwich supercell structure as in Fig. 2(a) above, which consists of a middle topological nontrivial ext-kagome-II, upper and lower sides trivial phase. However, unlike in ext-kagome-I, we cannot directly reverse the couplings K and J to achieve band inversion and ensure topological transitions. As a result, we need to find two different parameter configurations of ext-kagome-II that differ in topological phases and whose bulk bandgaps nearly coincide. We identify the different nontrivial or trivial topological phases by calculating $Z_2$ invariants under different parameters, and the detailed expression for the ext-kagome-II case is given in Section III. Then we scanned through the parameter space to select the good match for the bulk bandgaps. The pair of parameters we used are $K_1 =$

$-1.1$, $K_2 = -0.6$, $K_3 = -0.6$, $K_4 = -1.8$, $J = -1.9$ for nontrivial phase and $K_1 = -1.8$, $K_2 = -1.6$, $K_3 = -1.6$, $K_4 = -0.8$, $J = -0.6$ for trivial phase, and constructed the supercell as show in Fig. 4(a). Each layer consists of 12 unit cells. The edge states (blue line) associated with the pseudospin Hall effect are evident in Fig. 4(b). The mode profile of the edge states, which are localized at the interfaces, is exhibited in Fig. 4(c). The insets in Fig. 4(c) present the enlarged views of the boundaries of ext-kagome-II under different parameters.

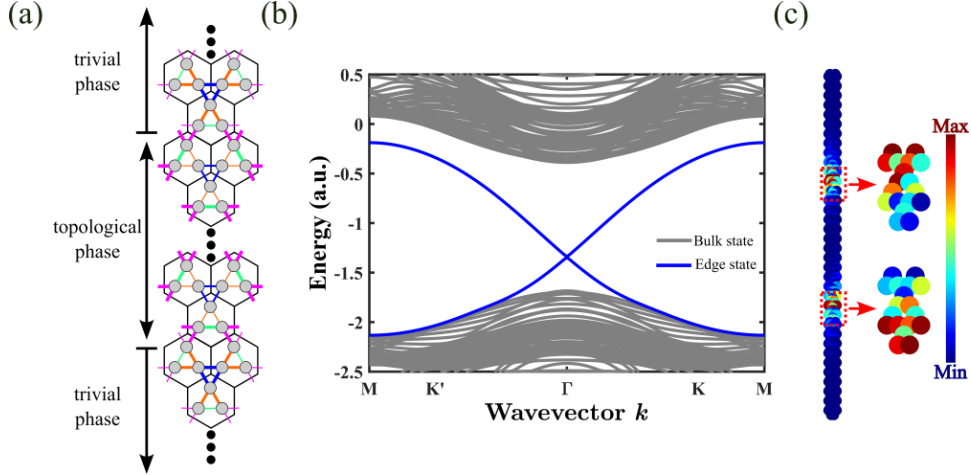

FIG. 4. Supercell and edge states of ext-kagome-II with domains. (a) Geometry demonstration of supercell composed of ext-kagome-II of different intracell coupling strengths. (b) The band structure of the supercell in (a). Gray lines represent bulk bands, and blue lines denote edge bands. (c) The mode profile of the edge state. Inset illustrations are the enlarged images of the edges at the interfaces.

To verify the helical nature and pseudospin selective excitation of the edge states in the extended kagome structures, we construct waveguides to analytically apply the excitations of opposite circular polarizations to observe the edge mode propagation. We introduced a helical excitation source at the middle of the domain wall within the TBM. The excitation is implemented by applying complex-valued on-site sources through the coupled-mode theory, where each source carries a phase term of $2n\pi/3$ (with $n = 0,1,2$). By tuning the phase sequence, left or right-handed helicity can be realized, which enables selective coupling to opposite pseudospin edge states. As shown in Fig. 5, all presented waveguides consist of a topological upper nontrivial domain and a lower trivial domain. Every domain is composed of 11 rows and 21 columns of unit cells. The excitation source is placed in the middle position of the waveguides. The source of the ext-kagome-I waveguide is applied to the lattice sites of the topological domain on the bottom side of the boundary. The circular arrows in the figures represent the excitation sources in the waveguides[5,68,69]. Note that in real systems like photonics, this color distribution of Bloch eigenvectors does not correspond to the full electromagnetic field strength in photonic systems, but is analogous to the intensity distribution of a single field component $E_z$, while not including the complete vectorial information ($E_x$, $E_y$), which is justified as topological classical wave applications are usually designed as single-mode on the resonators.

The frequencies of the source within the range of the edge states can excite the unidirectionally propagation. As shown in Figs. 5(a) and 5(b), the mode propagates along the left and right directions, respectively, and the amplitude gradually attenuates, which is introduced by the intrinsic loss term in the Hamiltonian to avoid the mode circling back to the source. The excitation with opposite polarization propagates in the opposite direction, which demonstrates ext-kagome-I possessing the pseudospin Hall effect like the Wu-Hu structure[12]. The waveguide of ext-kagome-II is similar to the waveguide of ext-kagome-I. As shown in Figs. 5 (c) and 5(d), when we apply the source with opposite polarization, the propagation direction of light changes. Similarly, the propagation changing with the light polarization reversing demonstrates ext-kagome-II also having the quantum pseudospin Hall effect and helical edge states. A slight back propagation is observed at the probe located five unit cells away from the source, with backscatter ratio (measured by flux) of approximately 8% (ext-kagome-I) and 1.6% (ext-kagome-II) of backscattering, which might be the result of imperfect edge modes due to the edge bandgaps, leading to scattering between pseudospin channels. In addition to propagating characteristics, the band inversion of double Dirac cones also shows very similar profiles of Wu-Hu structures. Near the Gamma point, the dipole modes $p\pm$ and quadrupole modes $d\pm$ lie on different doubly degenerate bands under trivial and topological cases(See Section III of Supplement Material[67] and Fig. S5), which is a typical characterization of pseudospin Hall systems.

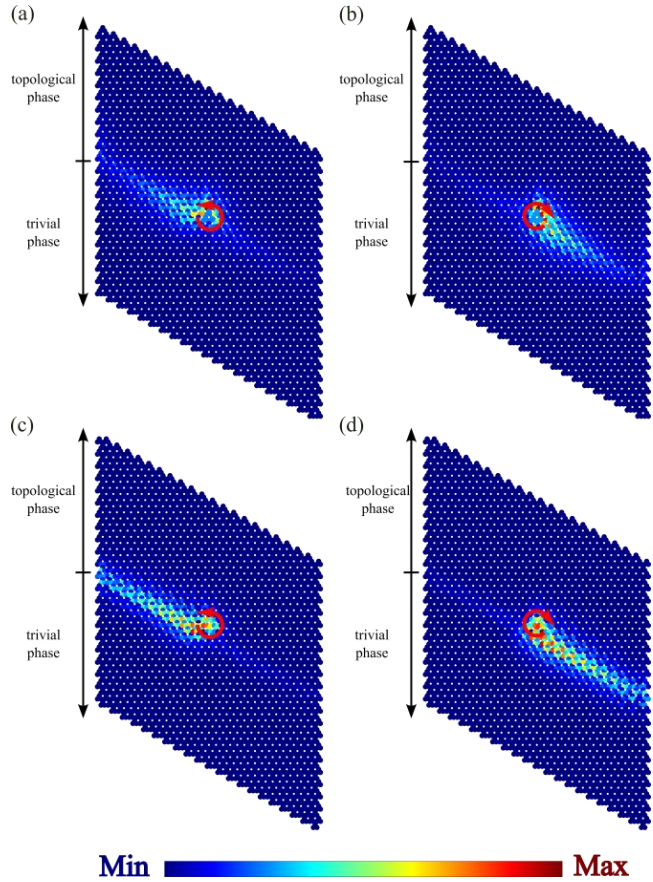

FIG. 5. The spin excitation of the extended kagome. (a)The spin excitation of ext-kagome-I. (b)The excitation with opposite-spin to (a) of ext-kagome-I. (c)The spin excitation of ext-kagome-II. (d) The excitation with opposite-spin to (c) of ext-kagome-II. Color bar in (a-d) shows the amplitude of the mode of excitation. The circular arrows denote the chiral excitation sources, with the arrows indicating the handedness (left-handed or right-handed) of the circular polarizations.

## B. Corner States in Extended Kagome Lattices

Conventional breathing kagome lattices are known to hold Wannier-type higher-order topology, where type-I corner states, type-II corner states, and more are discovered. To explore the existence of corner modes in extended kagome lattices, we looked into finite triangular structures and their mode distributions. First, we analytically constructed the array of finite triangular structures of the ext-kagome-I for observing the corner states. Each side of the finite triangular structure is composed of 12 topological unit cells with parameter (J/K = 5), and without any long-range couplings.

The energy spectrum of this finite triangular structure in Fig. 6(a) depicts the existence of the corner states. Figs. 6(c) and 6(d) are the enlarged eigenfrequency images of the red dashed box and green dashed box in Fig. 6(a), respectively. Interestingly, not only multiple groups of the type-I corner states (red dots) are observed, but also type-II corner states (green dots), which are not localized at the outermost corner, but near the two adjacent unit cells next to the outermost ones. Previous research [47,48] has shown that conventional kagome possesses one group of type-I corner states and two groups of type-II corner states under long-range interactions. Here, our system simultaneously hosts both type-I corner states and type-II corner states in the absence of long-range interactions. From the energy spectrum, the structure of ext-kagome-I supports 4 sets of type-I corner states with different frequencies inside different bandgaps, and 2 groups of type-II corner states. All groups of corner modes demonstrated here clearly lie in the bandgap, and type-II corner states have a similar profile of emerging from the edge spectrum, like in conventional kagome systems. For comparison, we constructed the same finite triangular structure using the trivial phase of the reverse parameters (K/J = 5) in Fig. 6(b) that doesn't support any corner states or edge states. The edge states of different frequencies in Figs. 6(c) and 6(d) (See Section V of Supplement Material[67]) are similar to the conventional kagome structure. Note that these finite structures do not have inversion symmetries due to the periodic directions and geometry of unit cells, which induce different corner mode distributions compared to conventional breathing kagome lattices.

Next, we demonstrate the mode profiles of type-I corner states in Figs. 6(e-h) and type-II corner states in Figs. 6(i-j) corresponding to different frequencies. From the mode profiles, we can observe that all these modes have good localizations and are triply degenerated due to $C_3$ symmetry. Interestingly, some type-I corner states are highly localized to the top corners of unit cells at the corner, and the type-I corner states in Fig. 6(f) are localized to the dimer beside the corner. Figs. 6(i-j) demonstrates one type of the type-II corner states, which occurs beside the unit cell at the corner. All the inset illustrations are the enlarged mode profiles of corner states for clarity.

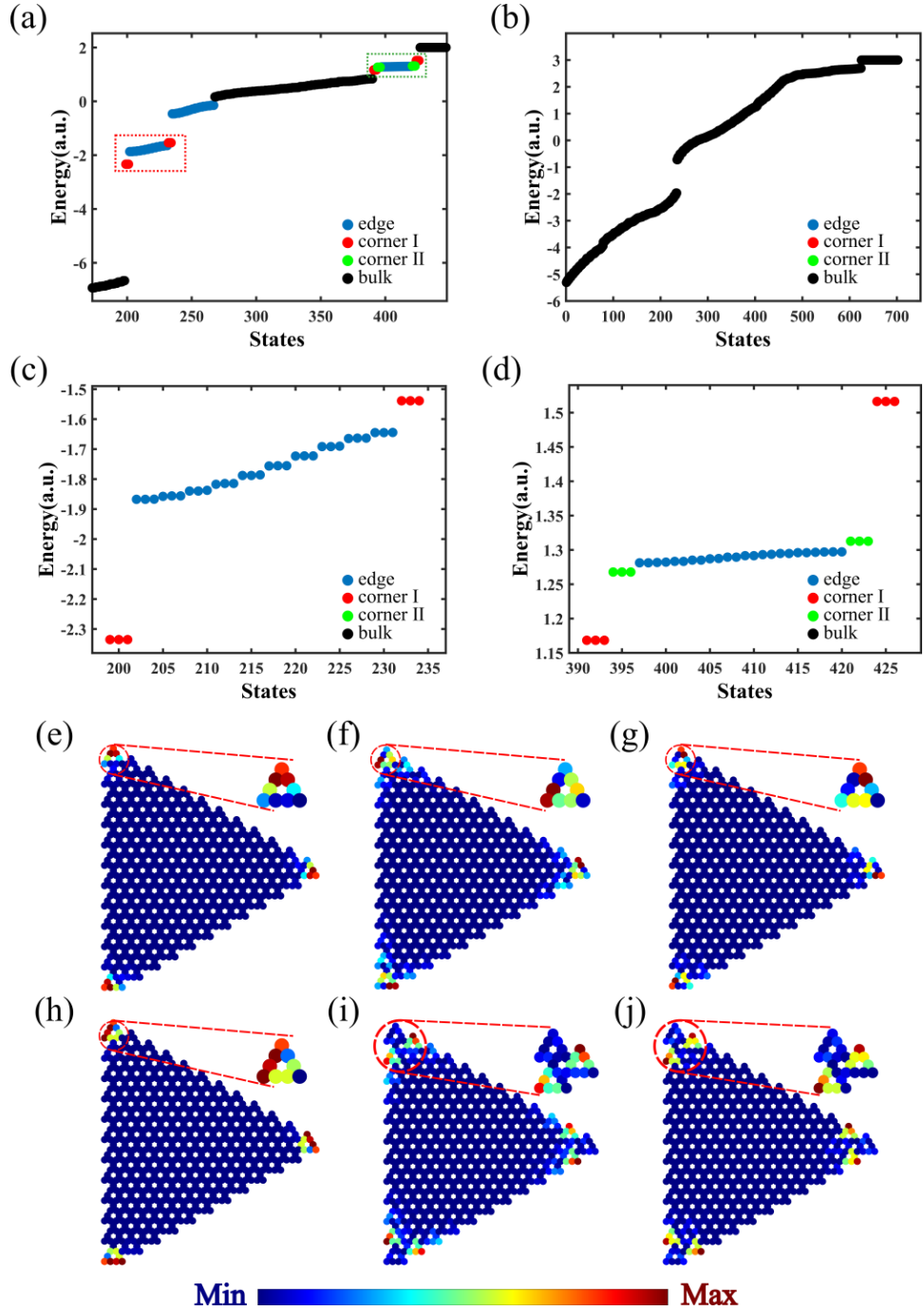

FIG. 6. The finite structure of ext-kagome-I. (a) The energy spectrum of ext-kagome-I triangular array for J/K = 5. Black, blue, red, and green dots represent the bulk states, edge states, type-I corner states, and type-II corner states, respectively. (b) The energy spectrum of a triangular-shaped array composed of the trivial phase for K/J = 5. (c) Enlarged image in red dashed box in (a). (d) The enlarged image in the green dashed box in (a). (e-i) Mode profiles of the type-I corner states with frequencies from low to high corresponding to the (a). (i-j) Mode profiles of type-II corner states with frequencies from low to high. The color bar shows the amplitude of the mode profile.

To demonstrate the existence of the corner states of ext-kagome-II, we also constructed a finite triangular-shaped ext-kagome-II structure with each edge composed of 12 unit cells with a group of topological domain parameters($K_1 = -2.0$, $K_2 = -1.2$, $K_3 = -1.2$, $K_4 = -0.8$, $J = -1.5$). The energy spectrum of the finite triangular structure in Fig. 7(a) demonstrates the existence of corner states (red dots). As shown in Fig. 7(b), the corner states at lower frequency emerge between two gapped edge states. As shown in Fig. 7(c), the corner states at higher frequency emerge between bulk states and edge states. Then we demonstrate the mode profiles of corner states corresponding to different frequencies in Fig. 7(d) and Fig. 7(e). The energy of lower-frequency corner states is focused on the outermost two sites. The maximum energy of the higher frequency corner states lies on one of the inner dots of the corner unit cell. Moreover, we find the corner states whose mode profiles are slightly less confined because the bandgap between the edge and bulk is small. We have also examined that multiple different groups of parameters are considered topological by the emergence of edge state band structure, holding similar corner states.

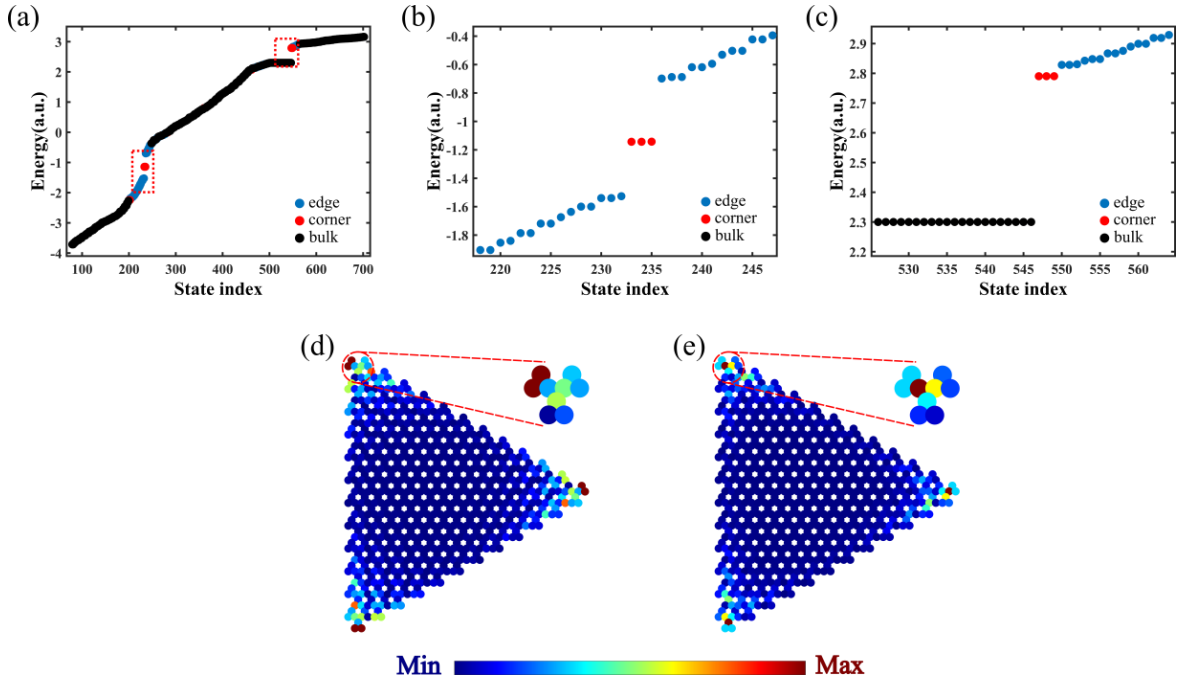

FIG. 7. The finite structure of ext-kagome-II. (a) The energy spectrum of ext-kagome-II triangular-shaped array for $K_1 = -2.0$, $K_2 = -1.2$, $K_3 = -1.2$, $K_4 = -0.8$, $J = -1.5$. Black, blue, and red dots represent the bulk states, edge states, and corner states, respectively. (b) The enlarged image in the lower red dashed box in (a). (c) The enlarged image in the higher red dashed box in (a). (d) Mode profiles of the corner states corresponding to the lower frequency in (b). (e) Mode profiles of the corner state corresponding to higher frequency in (c). The color bar shows the amplitude of the mode profile in (d-e).

On the relation between the emergence of corner states and higher-order topology, we need to clarify that the higher-order topology properties in extended kagome lattices are unknown, despite showing very similar properties with HOTI corner modes, including band structures, mode distributions, and robustness to disorders (Supplement VI). While we conducted Wannier center and bulk polarization calculations (Supplement IV), and interestingly, a large bulk polarization indicates bulk-edge-corner correspondence as far as we investigated, a quantized bulk polarization needed for HOTI classification cannot be steadily found and shows parameter dependence in both cases. This is possibly due to the extended kagome systems' unique multiple groups of corner states, even in a single bulk bandgap, which no current theory framework could investigate the bulk properties or the topological origin of the different corner states separately. We would also like to stress that, based on the theoretical tools currently available to us, we do not yet have a complete understanding of the physical origin of these non-quantized Wannier-center configurations. While the two-parameter regimes exhibit clearly distinguishable Wannier-geometry patterns, these differences do not correspond to any established quantized topological invariant or symmetry-based classification known in the literature. Developing a theoretical framework that can systematically interpret such non-quantized Wannier structures and multi-frequency corner states requires further investigation, which we regard as an interesting direction for future work.

### III. ANALYTICS

We employ the TBM to investigate the system with the coupling only between nearest-neighbor lattice sites. We derived an effective $4 \times 4$ Hamiltonian and calculated the $Z_2$ topological invariant for the ext-kagome-I system.

The base vectors are $\vec{a}_1 = a(0, 1)$ and $\vec{a}_2 = a(\sqrt{3}/2, 1/2)$ [Fig. S1], where $a$ is the lattice constant of ext-kagome-I unit cell. In the TBM, the system can approximately be described by Schrödinger equation as follow:

$$\widehat{H}|\psi\rangle = E|\psi\rangle \tag{2}$$

where the wave function under Wannier basis is expressed as:

$$|\psi\rangle = (\alpha_1, \alpha_2, \alpha_3, \alpha_4, \alpha_5, \alpha_6, \alpha_7, \alpha_8, \alpha_9)^T \tag{3}$$

and the Hamiltonian for the system under TBM is equal to

$$H_1 = \begin{pmatrix} 0 & K & K & 0 & 0 & Je^{-iak_y} & 0 & Je^{-iak_y} & 0 \\ K & 0 & K & K & 0 & 0 & 0 & 0 & Je^{i\left(\frac{1}{2}\sqrt{3}ak_x - \frac{ak_y}{2}\right)} \\ K & K & 0 & 0 & Je^{i\left(-\frac{1}{2}\sqrt{3}ak_x - \frac{ak_y}{2}\right)} & 0 & K & 0 & 0 \\ 0 & 0 & 0 & 0 & K & K & 0 & 0 & Je^{i\left(\frac{1}{2}\sqrt{3}ak_x - \frac{ak_y}{2}\right)} \\ 0 & 0 & Je^{i\left(\frac{1}{2}\sqrt{3}ak_x + \frac{ak_y}{2}\right)} & K & 0 & K & Je^{i\left(\frac{1}{2}\sqrt{3}ak_x + \frac{ak_y}{2}\right)} & 0 & 0 \\ Je^{iak_y} & 0 & 0 & K & K & 0 & 0 & K & 0 \\ 0 & 0 & K & 0 & Je^{i\left(-\frac{1}{2}\sqrt{3}ak_x - \frac{ak_y}{2}\right)} & 0 & 0 & K & K \\ Je^{iak_y} & 0 & 0 & 0 & 0 & K & K & 0 & K \\ 0 & Je^{i\left(-\frac{1}{2}\sqrt{3}ak_x + \frac{ak_y}{2}\right)} & 0 & Je^{i\left(-\frac{1}{2}\sqrt{3}ak_x + \frac{ak_y}{2}\right)} & 0 & 0 & K & K & 0 \end{pmatrix} \tag{4}$$

We are aiming to obtain the effective Hamiltonian of the 4 bands of doubly degenerated Dirac cones near the $\Gamma$ point. Other states far from the vicinity of the Dirac point should be ignored, but their interaction effects with the bands we study need to be considered in the effective Hamiltonian.

Utilizing the representation transformation of unitary matrix $U_1$ transform the TBM Hamiltonian into energy representation of $H_{01}$.

$$H^u = U_1^\dagger H U_1 = U_1^\dagger (H_{01} + H_{p1}) U_1 = H_{01}^u + H_{p1}^u \tag{5}$$

$$H_{01}^u = \begin{pmatrix} 4J & 0 & 0 & 0 & 0 & 0 & 0 & 0 & 0 \\ 0 & -2J & 0 & 0 & 0 & 0 & 0 & 0 & 0 \\ 0 & 0 & -2J & 0 & 0 & 0 & 0 & 0 & 0 \\ 0 & 0 & 0 & -2J & 0 & 0 & 0 & 0 & 0 \\ 0 & 0 & 0 & 0 & -2J & 0 & 0 & 0 & 0 \\ 0 & 0 & 0 & 0 & 0 & J & 0 & 0 & 0 \\ 0 & 0 & 0 & 0 & 0 & 0 & J & 0 & 0 \\ 0 & 0 & 0 & 0 & 0 & 0 & 0 & J & 0 \\ 0 & 0 & 0 & 0 & 0 & 0 & 0 & 0 & J \end{pmatrix} \qquad (6)$$

Define the subspace projection operator $P_{sub}$ of four degenerating energy bands near the vicinity of the Dirac point and the projection operator $P_{sub}^{\perp}$ of its orthogonal complement.

Decomposing $H_{p1}^u$ into component of orthogonal complement of subspace $H_{bd}^u$ and the component of interaction with subspace and its orthogonal complement $H_{bo}^u$.

$$H_{p1}^u = H_{bd}^u + H_{bo}^u \qquad (7)$$
$$H_{bd}^u = P_{sub} H_{p1}^u P_{sub} + P_{sub}^{\perp} H_{p1}^u P_{sub}^{\perp} \qquad (8)$$
$$H_{bo}^u = P_{sub} H_{p1}^u P_{sub}^{\perp} + P_{sub}^{\perp} H_{p1}^u P_{sub} \qquad (9)$$

We utilized the Schrieffer-Wolff transformation[63] to calculate $H_{e1}$:

$$H_{e1} = H_{01}^u + H_{bd}^u + \frac{1}{2}[S, H_{bo}^u] \qquad (10)$$

anti-Hermitian operator S satisfies:

$$H_{bo}^u + [S, H_{01}^u] = 0 \qquad (11)$$

Expanding $H_{e1}$ as quadratic term about $k_x$、$k_y$ around $\Gamma$ point, and keep linear term about $(K - J)$ assuming small perturbations when $K = J + o(K - J)$:

$$H_{e1} = \begin{pmatrix} \theta_1(k) + \varepsilon(k) & v_1(-k_x - ik_y) + \alpha(k_x - ik_y)^2 & \gamma(k_x - ik_y)^2 & \eta(k_x^2 + k_y^2) + \tau \\ v_1(-k_x + ik_y) + \alpha(k_x + ik_y)^2 & -\theta_1(k) + \varepsilon(k) & \eta(k_x^2 + k_y^2) + \tau & \kappa(k_x + ik_y)^2 \\ \gamma(k_x + ik_y)^2 & \eta(k_x^2 + k_y^2) + \tau & \theta_1(k) + \varepsilon(k) & v_1(k_x - ik_y) + \alpha(k_x + ik_y)^2 \\ \eta(k_x^2 + k_y^2) + \tau & \gamma(k_x - ik_y)^2 & v_1(k_x + ik_y) + \alpha(k_x - ik_y)^2 & -\theta_1(k) + \varepsilon(k) \end{pmatrix} \qquad (12)$$

where $\varepsilon(k) = (K - J)k^2/54 + (2K + J)/3$, $\theta_1(k) = \mu_1 + \beta_1 k^2$, $\mu_1 = K - J$, $\beta_1 = K/12$, $v_1 = (2K + 7J)/18$, $\alpha = (4K + 5J)/216$, $\gamma = 7(J - K)/108$, $\eta = -(4K + 5J)/108$, $\kappa = 5(K - J)/108$, $\tau = (J - K)/3$.

In the process of calculating the topological invariant, we ignore the identity matrix $H_I$ located on the diagonal, matrix located on the non-diagonal term $H_{bo}$, and keep terms of $H_\alpha$ to the first order of $k$. Thus, effective Hamiltonian is decomposed into two block-diagonal parts, and effective Hamiltonian have the similar form to Bernevig-Hughes-Zhang model[36].

$$H_1 = \begin{pmatrix} H_+ & 0 \\ 0 & H_- \end{pmatrix} = H_{BHZ} \qquad (13)$$

$$H_{\pm}^1 = \begin{pmatrix} \mu_I + \beta_I k^2 & v_I(-k_x \mp ik_y) \\ v_I(-k_x \pm ik_y) & -\mu_I - \beta_I k^2 \end{pmatrix} \qquad (14)$$

The form of the $H_{\pm}$ is the same as Dirac Hamiltonian, so there is an analytical solution for spin Chern number of the system[37]. Lastly, we obtain the $Z_2$ invariant of this system as

$$C_1 = \frac{1}{2}(sgn(\mu_1) - sgn(\beta_1)) = \frac{1}{2}(sgn(K - J) + 1) \qquad (15)$$

When considering cases of shrunken unit cells ($K < J < 0$), the systems are topologically trivial. in the cases of expanded ($J < K < 0$), the systems are topological.

For ext-kagome-II structure, we adopt the similar simplification methods to calculating the $Z_2$ invariant, Thus, effective Hamiltonian is decomposed into two block-diagonal parts, and effective Hamiltonian have the similar form to Bernevig-Hughes-Zhang model. The form of the $H_\pm^2$ is the same as Drica Hamiltonian.

$$H_\pm^2 = \begin{pmatrix} \mu_2 + \beta_2 k^2 & v_2(-k_x \mp ik_y) \\ v_2(-k_x \pm ik_y) & -\mu_2 - \beta_2 k^2 \end{pmatrix} \quad (16)$$

the topological invariant of this system as follow:

$$C_2 = \tfrac{1}{2}\big(sgn\,(\mu_2) - sgn(\beta_2)\big) \quad (17)$$

$\mu_2 = [2J^2 - 6JK_3 - 6JK_4 - K_1(4J + 3K_2 + 3K_3 + 3K_4) + K_2(-6J + 5K_3 + 5K_4) + 2K_1^2 + 4K_2^2 + 4K_3^2 + 4K_4^2 + 5K_3K_4]/[6(K_2 + K_3 + K_4)]$ (18)

$$\beta_2 = J[-6J/(2J + K_1) - 2J/(K_2 + K_3 + K_4) + 2K_1/(K_2 + K_3 + K_4) + 3]/12 \quad (19)$$

The $H_{e2}$ expanded to the quadratic term and linear term at the $\Gamma$ point and $K_i = J$, but there is a large freedom of choice of multiple intracell coupling parameters $K_i$ instead of the uniformly applied intracell coupling $K$ in the ext-kagome-I. Note that only when the condition for all intracell coupling coefficients $K_i = J + o(K_i - J)$ meets, the approximation is valid.

## IV. CONCLUSIONS

In this work, we proposed and constructed a new platform of pseudospin Dirac physics in two types of kagome-based structural systems, by arranging the unit cell selection methods and coupling tuning, different topological properties and localized mode distributions are shown. These new designs are achievable under current fabrication techniques similar to other current structures for topological photonics or other classical wave systems, and possibly realizable in photonics, acoustics, mechanics, and more. It is shown that type II corner states can also be achieved in systems without long-ranged coupling systems or local perturbations at the corners, enabling more ways to achieve multiple types of corner localized modes. Being able to parameter shift into kagome lattices and breathing kagome lattices, such systems also enable more types of topological waveguiding and trapping techniques, such as larger numbers of corner states, adiabatic Dirac waveguide, Dirac Vortex modes[70,71], and so on. An isolated gapped flat band in the systems increases the density of states and suppresses group velocity, which can enhance nonlinear interactions and support compact localized modes. With suitable nonlinearity and slight dispersion, the same platform can realize slow light and self-trapped or soliton-like states, coexisting with pseudospin Hall effects. Interestingly, these results also lead to the possibility that systems with $C_{3v}$ symmetry instead of $C_{6v}$ can also achieve good pseudospin Hall edge states and corner states. Noted that this system works especially well with systems where couplings are localized, such as surface plasmonic microwave and lower frequency regimes, acoustic cavities, and electronics, opens the possibility for applications such as multifrequency waveguides and cavities in EM waves, sensing, acoustic pseudospin devices, and MRI enhancers, while the properties are still showing some robustness under perturbations. Thus, this platform opens new possibilities to design and fabricate tunable and integrated devices, such as the topological pseudospin-controlled waveguides with more design freedom, and multi-frequency topological lasers.

**Competing interests**

The authors declare no conflict of interest.

**Acknowledgments**
M. L. acknowledges valuable discussions with Alexander Khanikaev. This project acknowledges financial support from the National Key R&D Program of China (Grant No. 2023YFB3811400), Shenzhen Science and Technology Innovation Commission (Grant No.20231204145504001), National Natural Science Foundation of China (Grant No.12404435), the Guangdong Science and Technology Commission (Grant No.2025A1515010774), and Tsinghua SIGS Scientific Research Startup Funding (Grant No. QD2023004C)

**Supplementary information**
Additional supporting information can be found online in the Supporting Information section at the end of this article.

# Topological Pseudospin Hall Effect and Multi-frequency Corner Modes in Kagome-based Lattices

Shenglong Guo[1*], Qinhui Jiang[1*], Yuma Kawaguchi[2], Bo Li[1,3], and Mengyao Li[1†]

[1]Tsinghua Shenzhen International Graduate School, Tsinghua University, Shenzhen 518055, China

[2]Department of Electrical Engineering, The City College of New York, New York, NY 10031, USA

[3]Suzhou Laboratory, Suzhou 215000, China

*These authors contributed equally.

†Contact information: mengyaoli@sz.tsinghua.edu.cn

**Supplemental Text**
I. Supplemental Derivation of the Effective Hamiltonian and Topological Invariant
    A. The first type of extended kagome (ext-kagome-I)
    B. The second type of extended kagome (ext-kagome-II)
    C. γ-matrix decomposition of the effective Hamiltonian for the ext-kagome-I
II. The parameter space of the ext-kagome-II
III. Mode profiles evolution with band inversion
IV. The Wannier bands of the extended kagome
V. The mode profiles of the edge state at the different frequencies
VI. The random perturbation of corner states

# I. Supplemental Derivation of the Effective Hamiltonian and Topological Invariant
## A. The first type of extended kagome (ext-kagome-I)

We construct the ext-kagome-I lattice using the lattice vectors $\vec{a}_1 = a\,(0,1)$ and $\vec{a}_2 = a\,(\sqrt{3}/2, 1/2)$ [Fig. S1], where $a$ is the lattice constant of ext-kagome-I unit cell.

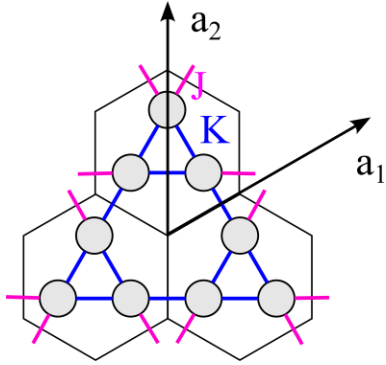

FIG. S1. The structure of the ext-kagome-I unit cell.

We decomposed the TBM Hamiltonian $H_1$ of the ext-kagome-I into $H_{01}$ representing the Hamiltonian at the $\Gamma$ point ($k=0$) and the perturbation term $H_{p1}$ when $K = J$:

$$H_1 = H_{01} + H_{p1} \tag{S1}$$

$$H_{01} = \begin{pmatrix} 0 & J & J & 0 & 0 & J & 0 & J & 0 \\ J & 0 & J & J & 0 & 0 & 0 & 0 & J \\ J & J & 0 & 0 & J & 0 & J & 0 & 0 \\ 0 & J & 0 & 0 & J & J & 0 & 0 & J \\ 0 & 0 & J & J & 0 & J & J & 0 & 0 \\ J & 0 & 0 & J & J & 0 & 0 & J & 0 \\ 0 & 0 & J & 0 & J & 0 & 0 & J & J \\ J & 0 & 0 & 0 & 0 & J & J & 0 & J \\ 0 & J & 0 & J & 0 & 0 & J & J & 0 \end{pmatrix} \tag{S2}$$

$$H_{p1} = \begin{pmatrix} 0 & K-J & K-J & 0 & 0 & -J+Je^{-iak_y} & 0 & -J+Je^{-iak_y} & 0 \\ K-J & 0 & K-J & K-J & 0 & 0 & 0 & 0 & -J+Je^{i\left(\frac{\sqrt{3}}{2}ak_x-\frac{1}{2}ak_y\right)} \\ K-J & K-J & 0 & 0 & -J+Je^{i\left(-\frac{\sqrt{3}}{2}ak_x-\frac{1}{2}ak_y\right)} & 0 & K-J & 0 & 0 \\ 0 & 0 & 0 & 0 & K-J & K-J & 0 & 0 & -J+Je^{i\left(\frac{\sqrt{3}}{2}ak_x-\frac{1}{2}ak_y\right)} \\ 0 & 0 & -J+Je^{i\left(\frac{\sqrt{3}}{2}ak_x+\frac{1}{2}ak_y\right)} & K-J & 0 & K-J & -J+Je^{i\left(\frac{\sqrt{3}}{2}ak_x+\frac{1}{2}ak_y\right)} & 0 & 0 \\ -J+Je^{iak_y} & 0 & 0 & K-J & K-J & 0 & 0 & K-J & 0 \\ 0 & 0 & K-J & 0 & -J+Je^{i\left(-\frac{\sqrt{3}}{2}ak_x-\frac{1}{2}ak_y\right)} & 0 & 0 & K-J & K-J \\ -J+Je^{iak_y} & 0 & 0 & 0 & 0 & K-J & K-J & 0 & K-J \\ 0 & -J+Je^{i\left(\frac{\sqrt{3}}{2}ak_x+\frac{1}{2}ak_y\right)} & 0 & -J+Je^{i\left(-\frac{\sqrt{3}}{2}ak_x+\frac{1}{2}ak_y\right)} & 0 & 0 & K-J & K-J & 0 \end{pmatrix} \tag{S3}$$

The effective Hamiltonian is determined by the base vectors in the Hilbert space. We choose the base vectors formed by the eigenmodes when the coupling strength of the intercell is equal to that of the intracell. Those eigenmodes represent eigenvectors of the TBM Hamiltonian at the $\Gamma$ point when $K = J$, and are composed into a unitary matrix $U_1$ as follow[1]:

$$|u_1\rangle_1 = \left(\frac{1}{3}, \frac{1}{3}, \frac{1}{3}, \frac{1}{3}, \frac{1}{3}, \frac{1}{3}, \frac{1}{3}, \frac{1}{3}, \frac{1}{3}\right)^T \tag{S4}$$

$$|u_2\rangle_1 = \left(\frac{1}{3}, \frac{1}{3}e^{\frac{2i\pi}{3}}, \frac{1}{3}e^{-\frac{2i\pi}{3}}, \frac{1}{3}, \frac{1}{3}e^{\frac{2i\pi}{3}}, \frac{1}{3}e^{-\frac{2i\pi}{3}}, \frac{1}{3}, \frac{1}{3}e^{\frac{2i\pi}{3}}, \frac{1}{3}e^{-\frac{2i\pi}{3}}\right)^T \tag{S5}$$

$$|u_3\rangle_1 = \left(\frac{1}{3}, \frac{1}{3}e^{-\frac{2i\pi}{3}}, \frac{1}{3}e^{\frac{2i\pi}{3}}, \frac{1}{3}, \frac{1}{3}e^{-\frac{2i\pi}{3}}, \frac{1}{3}e^{\frac{2i\pi}{3}}, \frac{1}{3}, \frac{1}{3}e^{-\frac{2i\pi}{3}}, \frac{1}{3}e^{\frac{2i\pi}{3}}\right)^T \tag{S6}$$

$$|u_4\rangle_1 = \left(\frac{1}{3}, \frac{1}{3}e^{-\frac{2i\pi}{3}}, \frac{1}{3}e^{\frac{2i\pi}{3}}, \frac{1}{3}e^{\frac{2i\pi}{3}}, \frac{1}{3}, \frac{1}{3}e^{-\frac{2i\pi}{3}}, \frac{1}{3}e^{-\frac{2i\pi}{3}}, \frac{1}{3}e^{\frac{2i\pi}{3}}, \frac{1}{3}\right)^T \tag{S7}$$

$$|u_5\rangle_1 = \left(\frac{1}{3}, \frac{1}{3}e^{\frac{2i\pi}{3}}, \frac{1}{3}e^{-\frac{2i\pi}{3}}, \frac{1}{3}e^{-\frac{2i\pi}{3}}, \frac{1}{3}, \frac{1}{3}e^{\frac{2i\pi}{3}}, \frac{1}{3}e^{\frac{2i\pi}{3}}, \frac{1}{3}e^{-\frac{2i\pi}{3}}, \frac{1}{3}\right)^T \tag{S8}$$

$$|u_6\rangle_1 = \left(\frac{1}{3}, \frac{1}{3}, \frac{1}{3}, \frac{1}{3}e^{\frac{2i\pi}{3}}, \frac{1}{3}e^{\frac{2i\pi}{3}}, \frac{1}{3}e^{\frac{2i\pi}{3}}, \frac{1}{3}e^{-\frac{2i\pi}{3}}, \frac{1}{3}e^{-\frac{2i\pi}{3}}, \frac{1}{3}e^{-\frac{2i\pi}{3}}\right)^T \tag{S9}$$

$$|u_7\rangle_1 = \left(\frac{1}{3}, \frac{1}{3}e^{\frac{2i\pi}{3}}, \frac{1}{3}e^{-\frac{2i\pi}{3}}, \frac{1}{3}e^{\frac{2i\pi}{3}}, \frac{1}{3}e^{-\frac{2i\pi}{3}}, \frac{1}{3}e^{-\frac{2i\pi}{3}}, \frac{1}{3}, \frac{1}{3}, \frac{1}{3}e^{\frac{2i\pi}{3}}\right)^T \tag{S10}$$

$$|u_8\rangle_1 = \left(\frac{1}{3}, \frac{1}{3}, \frac{1}{3}, \frac{1}{3}e^{-\frac{2i\pi}{3}}, \frac{1}{3}e^{-\frac{2i\pi}{3}}, \frac{1}{3}e^{-\frac{2i\pi}{3}}, \frac{1}{3}e^{\frac{2i\pi}{3}}, \frac{1}{3}e^{\frac{2i\pi}{3}}, \frac{1}{3}e^{\frac{2i\pi}{3}}\right)^T \tag{S11}$$

$$|u_9\rangle_1 = \left(\frac{1}{3}, \frac{1}{3}e^{-\frac{2i\pi}{3}}, \frac{1}{3}e^{\frac{2i\pi}{3}}, \frac{1}{3}e^{-\frac{2i\pi}{3}}, \frac{1}{3}e^{\frac{2i\pi}{3}}, \frac{1}{3}, \frac{1}{3}e^{\frac{2i\pi}{3}}, \frac{1}{3}, \frac{1}{3}e^{-\frac{2i\pi}{3}}\right)^T \tag{S12}$$

$$U_1 = (|u_1\rangle, |u_2\rangle, |u_3\rangle, |u_4\rangle, |u_5\rangle, |u_6\rangle, |u_7\rangle, |u_8\rangle, |u_9\rangle) \tag{S13}$$

$|u_1\rangle_1$ describes the non-degenerate state far from the vicinity of the Dirac point and at the bottom of the energy spectrum. $|u_2\rangle_1, |u_3\rangle_1, |u_4\rangle_1, |u_5\rangle_1$ describe the two non-degenerate states and two doubly degenerate states degenerating far from the vicinity of the Dirac point and at the top of the energy spectrum. $|u_6\rangle_1, |u_7\rangle_1, |u_8\rangle_1, |u_9\rangle_1$ describe the four doubly degenerate states degenerating near the vicinity of the Dirac point. Equations (S4-S12) characterize the energy bands at the Γ point for the system with $J = K$, corresponding to the band structure shown in Fig. 1(b). In this unperturbed case, the eigenstates $|u_6\rangle_1, |u_7\rangle_1, |u_8\rangle_1, |u_9\rangle_1$ describe the topmost four bands. When we introduce a perturbation such that $J = K + O(K)$ (leading to Fig. 1(c)), these states evolve adiabatically into the new eigenstates $|u_6'\rangle_1, |u_7'\rangle_1, |u_8'\rangle_1, |u_9'\rangle_1$, which we use to describe the four doubly degenerate states degenerating near the vicinity of the Dirac point.

We employ the Schrieffer–Wolff transformation to obtain the effective Hamiltonian, because a simple projection captures only the leading-order Hamiltonian within the selected subspace, while neglecting the coupling between the selected subspace and other subspaces. These second-order processes are essential in our system, particularly near parameter regimes with near degeneracies, as they generate the correct renormalized couplings and level shifts. The Schrieffer–Wolff approach systematically incorporates these effects, whereas projection alone would give an incomplete and generally inaccurate effective theory.

For clarity, we present the derivation following the standard Schrieffer–Wolff sequence: the Hamiltonian is first expanded using the Baker-Campbell-Hausdorff (BCH) formula, the Luttinger–Kohn condition is then imposed to remove the first-order off-block-diagonal terms, and the resulting expression is finally simplified to obtain Eq. (10). We note that Eq. (11) ensures the first-order block-off-diagonal terms are zero and ignores higher-order block-off-diagonal terms, as they only contribute to the higher-order Schrieffer-Wolff expansion and are irrelevant for determining the first-order generator. This is consistent with the standard lowest-order Luttinger–Kohn condition.

As shown in Fig. S2, the effective Hamiltonian $H_{e1}$ provides a good approximation when considering the system near the Γ point.

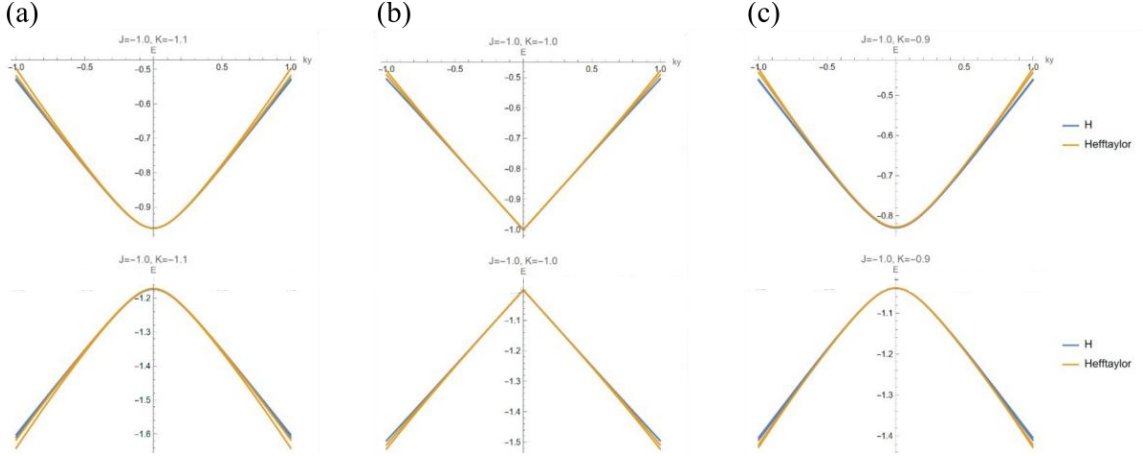

FIG. S2. The energy band dispersion calculated by $H_1$ and $H_{e1}$ with different coupling strengths. (a) $J = -1.0, K = -1.1$; (b) $J = -1.0, K = -1.0$; (c) $J = -1.0, K = -0.9$.

Then, we calculate the topological invariant of the effective Hamiltonian $H_{e1}$. Firstly, $H_{e1}$ is decomposed into parts as follows:

$$H_{e1} = H_{BHZ} + H_I + H_\alpha + H_{bo} \tag{S14}$$

$$H_{BHZ} = \begin{pmatrix} \theta_1(k) & v_1(-k_x - ik_y) & 0 & 0 \\ v_1(-k_x + ik_y) & -\theta_1(k) & 0 & 0 \\ 0 & 0 & \theta_1(k) & v_1(k_x - ik_y) \\ 0 & 0 & v_1(k_x + ik_y) & -\theta_1(k) \end{pmatrix} \tag{S15}$$

$$H_I = \begin{pmatrix} \varepsilon(k) & 0 & 0 & 0 \\ 0 & \varepsilon(k) & 0 & 0 \\ 0 & 0 & \varepsilon(k) & 0 \\ 0 & 0 & 0 & \varepsilon(k) \end{pmatrix} \tag{S16}$$

$$H_\alpha = \begin{pmatrix} 0 & \alpha(k_x - ik_y)^2 & 0 & 0 \\ \alpha(k_x + ik_y)^2 & 0 & 0 & 0 \\ 0 & 0 & 0 & \alpha(k_x + ik_y)^2 \\ 0 & 0 & \alpha(k_x - ik_y)^2 & 0 \end{pmatrix} \tag{S17}$$

$$H_{bo} = \begin{pmatrix} 0 & 0 & \gamma(k_x - ik_y)^2 & \eta(k_x^2 + k_y^2) + \tau \\ 0 & 0 & \eta(k_x^2 + k_y^2) + \tau & \kappa(k_x + ik_y)^2 \\ \gamma(k_x + ik_y)^2 & \eta(k_x^2 + k_y^2) + \tau & 0 & 0 \\ \eta(k_x^2 + k_y^2) + \tau & \kappa(k_x - ik_y)^2 & 0 & 0 \end{pmatrix} \tag{S18}$$

We note that the higher-order terms in Eqs. (S17) and (S18) do not modify the $Z_2$ invariant, as they vanish at the $\Gamma$ point and only contribute analytic $k^2$ corrections that cannot change the band inversion or the parity eigenvalues at TRIMs. Therefore, the topological classification can be consistently performed using the BHZ term Eq. (S15) alone.

We note that Eq. (15) corresponds to the pseudo-spin Chern number rather than the $Z_2$ index. In the present system, the Schrieffer–Wolff effective Hamiltonian is nearly block-diagonal, and the pseudo-spin sectors are only weakly mixed. In this limit, the pseudo-spin Chern number serves as a reliable numerical indicator of the non-trivial

$Z_2$ topology, and a non-zero value of Eq. (15) thus signals the $Z_2$-odd phase.

**B. The second type extended kagome (ext-kagome-II)**

In this case, we use the same method above to study the effective Hamiltonian and the topological invariant. The structure of ext-kagome-II is shown in Fig. S2.

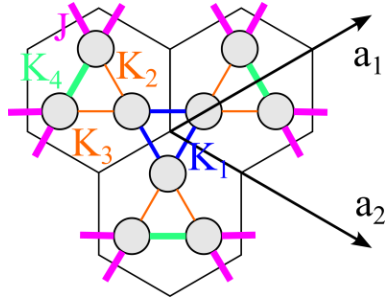

FIG. S3. The structure of the ext-kagome-II unit cell.

The base vectors are $\vec{a}_1 = a\,(\sqrt{3}/2, 1/2)$ and $\vec{a}_2 = a\,(\sqrt{3}/2, -1/2)$ [Fig. S3], where $a$ is the lattice constant. In the TBM, the Hamiltonian $H_2$ for the system is equal to

$$H_2 = \begin{pmatrix} 0 & 0 & K_4 & K_2 & 0 & Je^{i\left(\frac{1}{2}ak_y - \frac{\sqrt{3}}{2}ak_x\right)} & 0 & Je^{iak_y} & 0 \\ 0 & 0 & Je^{i\left(\frac{\sqrt{3}}{2}ak_x + \frac{1}{2}ak_y\right)} & 0 & K_3 & K_4 & 0 & 0 & Je^{iak_y} \\ K_4 & Je^{-i\left(\frac{\sqrt{3}}{2}ak_x + \frac{1}{2}ak_y\right)} & 0 & K_3 & 0 & 0 & 0 & 0 & Je^{i\left(\frac{1}{2}ak_y - \frac{\sqrt{3}}{2}ak_x\right)} \\ K_2 & 0 & K_3 & 0 & K_1 & 0 & K_1 & 0 & 0 \\ 0 & K_3 & 0 & K_1 & 0 & K_2 & K_1 & 0 & 0 \\ Je^{i\left(\frac{\sqrt{3}}{2}ak_x - \frac{1}{2}ak_y\right)} & K_4 & 0 & 0 & K_2 & 0 & 0 & Je^{i\left(\frac{\sqrt{3}}{2}ak_x + \frac{1}{2}ak_y\right)} & 0 \\ 0 & 0 & 0 & K_1 & K_1 & 0 & 0 & K_2 & K_3 \\ Je^{-iak_y} & 0 & 0 & 0 & 0 & Je^{-i\left(\frac{\sqrt{3}}{2}ak_x + \frac{1}{2}ak_y\right)} & K_2 & 0 & K_4 \\ 0 & Je^{-iak_y} & Je^{i\left(\frac{\sqrt{3}}{2}ak_x - \frac{1}{2}ak_y\right)} & 0 & 0 & 0 & K_3 & K_4 & 0 \end{pmatrix} \quad (S19)$$

Decomposing the TBM Hamiltonian $H_2$ into $H_{02}$ representing the Hamiltonian at the $\Gamma$ point $(k = 0)$ and the perturbation term $H_{p2}$ when $K_1 = K_2 = K_3 = K_4 = J$:

$$H_2 = H_{02} + H_{p2} \tag{S20}$$

$$H_{02} = \begin{pmatrix} 0 & 0 & J & J & 0 & J & 0 & J & 0 \\ 0 & 0 & J & 0 & J & J & 0 & 0 & J \\ J & J & 0 & J & 0 & 0 & 0 & 0 & J \\ J & 0 & J & 0 & J & 0 & J & 0 & 0 \\ 0 & J & 0 & J & 0 & J & J & 0 & 0 \\ J & J & 0 & 0 & J & 0 & 0 & J & 0 \\ 0 & 0 & 0 & J & J & 0 & 0 & J & J \\ J & 0 & 0 & 0 & 0 & J & J & 0 & J \\ 0 & J & J & 0 & 0 & 0 & J & J & 0 \end{pmatrix} \tag{S21}$$

$$H_{p2} = \begin{pmatrix} 0 & 0 & K_4-J & K_2-J & 0 & -J+Je^{i(\frac{1}{2}ak_y-\frac{\sqrt{3}}{2}ak_x)} & 0 & -J+Je^{iak_y} & 0 \\ 0 & 0 & -J+Je^{i(\frac{\sqrt{3}}{2}ak_x+\frac{1}{2}ak_y)} & 0 & -J+K_3 & -J+K_4 & 0 & 0 & -J+Je^{iak_y} \\ K_4-J & -J+Je^{-i(\frac{\sqrt{3}}{2}ak_x+\frac{1}{2}ak_y)} & 0 & K_3-J & 0 & 0 & 0 & 0 & -J+Je^{i(\frac{1}{2}ak_y-\frac{\sqrt{3}}{2}ak_x)} \\ K_2-J & 0 & K_3-J & 0 & K_1-J & 0 & K_1-J & 0 & 0 \\ 0 & K_3-J & 0 & K_1-J & 0 & K_2-J & K_1-J & 0 & 0 \\ -J+Je^{i(\frac{\sqrt{3}}{2}ak_x-\frac{1}{2}ak_y)} & K_4-J & 0 & 0 & K_2-J & 0 & 0 & -J+Je^{i(\frac{\sqrt{3}}{2}ak_x+\frac{1}{2}ak_y)} & 0 \\ 0 & 0 & 0 & K_1-J & K_1-J & 0 & 0 & K_2-J & K_3-J \\ -J+Je^{-iak_y} & 0 & 0 & 0 & 0 & -J+Je^{-i(\frac{\sqrt{3}}{2}ak_x+\frac{1}{2}ak_y)} & K_2-J & 0 & K_4-J \\ 0 & -J+Je^{-iak_y} & -J+Je^{i(\frac{\sqrt{3}}{2}ak_x-\frac{1}{2}ak_y)} & 0 & 0 & 0 & K_3-J & K_4-J & 0 \end{pmatrix} \quad (S22)$$

We choose the base vector formed by the eigenmodes when the coupling strength of the intercell is equal to that of the intracell, as in ext-kagome-I. The unitary matrix $U_2$ as follows:

$$|u_1\rangle_2 = \left(\frac{1}{3},\frac{1}{3},\frac{1}{3},\frac{1}{3},\frac{1}{3},\frac{1}{3},\frac{1}{3},\frac{1}{3},\frac{1}{3}\right)^T \tag{S23}$$

$$|u_2\rangle_2 = \left(\frac{1}{3},\frac{1}{3},\frac{1}{3}e^{\frac{2i\pi}{3}},\frac{1}{3}e^{-\frac{2i\pi}{3}},\frac{1}{3}e^{\frac{2i\pi}{3}},\frac{1}{3}e^{-\frac{2i\pi}{3}},\frac{1}{3},\frac{1}{3}e^{\frac{2i\pi}{3}},\frac{1}{3}e^{-\frac{2i\pi}{3}}\right)^T \tag{S24}$$

$$|u_3\rangle_2 = \left(\frac{1}{3},\frac{1}{3},\frac{1}{3}e^{-\frac{2i\pi}{3}},\frac{1}{3}e^{\frac{2i\pi}{3}},\frac{1}{3}e^{-\frac{2i\pi}{3}},\frac{1}{3}e^{\frac{2i\pi}{3}},\frac{1}{3},\frac{1}{3}e^{-\frac{2i\pi}{3}},\frac{1}{3}e^{\frac{2i\pi}{3}}\right)^T \tag{S25}$$

$$|u_4\rangle_2 = \left(\frac{1}{3},\frac{1}{3}e^{-\frac{2i\pi}{3}},\frac{1}{3}e^{\frac{2i\pi}{3}},\frac{1}{3}e^{-\frac{2i\pi}{3}},\frac{1}{3},\frac{1}{3}e^{\frac{2i\pi}{3}},\frac{1}{3}e^{\frac{2i\pi}{3}},\frac{1}{3}e^{-\frac{2i\pi}{3}},\frac{1}{3}\right)^T \tag{S26}$$

$$|u_5\rangle_2 = \left(\frac{1}{3},\frac{1}{3}e^{\frac{2i\pi}{3}},\frac{1}{3}e^{-\frac{2i\pi}{3}},\frac{1}{3}e^{\frac{2i\pi}{3}},\frac{1}{3},\frac{1}{3}e^{-\frac{2i\pi}{3}},\frac{1}{3}e^{-\frac{2i\pi}{3}},\frac{1}{3}e^{\frac{2i\pi}{3}},\frac{1}{3}\right)^T \tag{S27}$$

$$|u_6\rangle_2 = \left(\frac{1}{3},\frac{1}{3}e^{-\frac{2i\pi}{3}},\frac{1}{3},\frac{1}{3},\frac{1}{3}e^{-\frac{2i\pi}{3}},\frac{1}{3}e^{-\frac{2i\pi}{3}},\frac{1}{3}e^{\frac{2i\pi}{3}},\frac{1}{3}e^{\frac{2i\pi}{3}},\frac{1}{3}e^{\frac{2i\pi}{3}}\right)^T \tag{S28}$$

$$|u_7\rangle_2 = \left(\frac{1}{3},\frac{1}{3}e^{-\frac{2i\pi}{3}},\frac{1}{3}e^{-\frac{2i\pi}{3}},\frac{1}{3}e^{\frac{2i\pi}{3}},\frac{1}{3}e^{\frac{2i\pi}{3}},\frac{1}{3},\frac{1}{3}e^{\frac{2i\pi}{3}},\frac{1}{3},\frac{1}{3}e^{-\frac{2i\pi}{3}}\right)^T \tag{S29}$$

$$|u_8\rangle_2 = \left(\frac{1}{3},\frac{1}{3}e^{\frac{2i\pi}{3}},\frac{1}{3},\frac{1}{3},\frac{1}{3}e^{\frac{2i\pi}{3}},\frac{1}{3}e^{\frac{2i\pi}{3}},\frac{1}{3}e^{-\frac{2i\pi}{3}},\frac{1}{3}e^{-\frac{2i\pi}{3}},\frac{1}{3}e^{-\frac{2i\pi}{3}}\right)^T \tag{S30}$$

$$|u_9\rangle_2 = \left(\frac{1}{3},\frac{1}{3}e^{\frac{2i\pi}{3}},\frac{1}{3}e^{\frac{2i\pi}{3}},\frac{1}{3}e^{-\frac{2i\pi}{3}},\frac{1}{3}e^{-\frac{2i\pi}{3}},\frac{1}{3},\frac{1}{3}e^{-\frac{2i\pi}{3}},\frac{1}{3},\frac{1}{3}e^{\frac{2i\pi}{3}}\right)^T \tag{S31}$$

$$U_2 = (|u_1\rangle,|u_2\rangle,|u_3\rangle,|u_4\rangle,|u_5\rangle,|u_6\rangle,|u_7\rangle,|u_8\rangle,|u_9\rangle) \tag{S32}$$

$|u_6\rangle_2, |u_7\rangle_2, |u_8\rangle_2, |u_9\rangle_2$ describe the four doubly degenerate states degenerating near the vicinity of the Dirac point. $|u_1\rangle_2, |u_2\rangle_2, |u_3\rangle_2, |u_4\rangle_2, |u_5\rangle_2$ describe the five states far from the vicinity of the Dirac point.

We use the same method as the ext-kagome-I to obtain the similar $H_{e2}$ as $H_{e1}$ and expanding $H_{e2}$ as a quadratic term about $k_x$、$k_y$ and linear term about $(K_i - J)$ $(i = 1,2,3,4)$ at the $\Gamma$ point when $K_1 = K_2 = K_3 = K_4 = J$. The derivation of the topological invariant for this second structure follows analogously to the first. The off-diagonal $k^2$ terms are neglected based on the same perturbative argument as before, as they constitute higher-order contributions that do not affect the value of the $Z_2$ invariant. We simultaneously calculated the dispersion of the $H_2$ and $H_{e2}$. The energy bands of the $H_2$ and $H_{e2}$ are nearly identical in the vicinity of the $\Gamma$ point shown in Fig. S4.

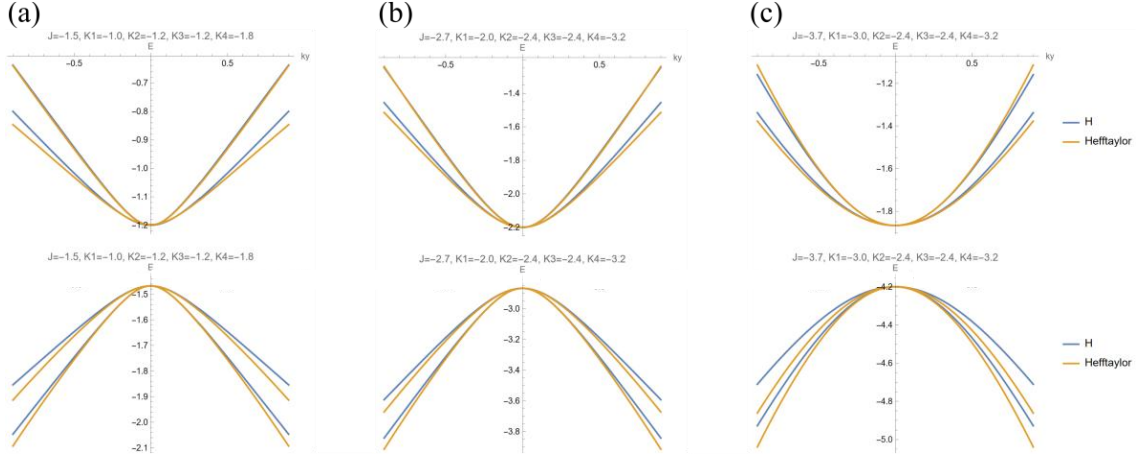

FIG. S4. The energy band dispersion calculated by $H_2$ and $H_2$ with different coupling strengths. (a) $J = -1.5, K_1 = -1.0, K_2 = -1.2, K_3 = -1.2, K_4 = -1.8$; (b) $J = -2.7, K_1 = -2.0, K_2 = -2.4, K_3 = -2.4, K_4 = -3.2$; (c) $J = -3.7, K_1 = -3.0, K_2 = -2.4, K_3 = -2.4, K_4 = -3.2$.

### C. γ-matrix decomposition of the effective Hamiltonian for the ext-kagome-I

To further confirm whether the type of our structure corresponds to the pseudospin-type or the valley-type, in this Supplement, we provide a full γ-matrix decomposition of the effective Hamiltonian for the ext-kagome-I model, extracting all expansion coefficients (denoted as R33–R53). This analysis confirms that the low-energy theory contains no valley-Hall mass component, leaving pseudospin physics as the sole mechanism for the gap opening and topology.

On our basis, $I_4$ is a scalar, $\gamma^0$–$\gamma^3$ form a vector set, and $\sigma^{uv}$ corresponds to a second-rank tensor. $\gamma^5$ is the pseudoscalar basis and $\gamma^5 \gamma^x (x = 0,1,2,3)$ is the axial vector basis. $S$ denotes the scalar sector, $V_0 - V_3$ the vector sector, $T$, the tensor sector, $P$, the pseudoscalar sector, and $A_0 - A_3$, the axial-vector sector.

$$H = S I_4 + V_3 \gamma^3 + \tfrac{1}{2} T_{01} \sigma^{01} + \tfrac{1}{2} T_{03} \sigma^{03} + \tfrac{1}{2} T_{12} \sigma^{12} + \tfrac{1}{2} T_{13} \sigma^{13} + \tfrac{1}{2} T_{23} \sigma^{23} + P \gamma^5 + A_0 \gamma^5 \gamma^0 + A_1 \gamma^5 \gamma^1 + A_2 \gamma^5 \gamma^2 \quad (S33)$$

$$I_4 = \begin{pmatrix} 1 & 0 & 0 & 0 \\ 0 & 1 & 0 & 0 \\ 0 & 0 & 1 & 0 \\ 0 & 0 & 0 & 1 \end{pmatrix} \quad (S34)$$

$$\gamma^0 = \begin{pmatrix} 1 & 0 & 0 & 0 \\ 0 & 1 & 0 & 0 \\ 0 & 0 & -1 & 0 \\ 0 & 0 & 0 & -1 \end{pmatrix} \quad (S35)$$

$$\gamma^1 = \begin{pmatrix} 0 & 0 & 0 & 1 \\ 0 & 0 & 1 & 0 \\ 0 & -1 & 0 & 0 \\ -1 & 0 & 0 & 0 \end{pmatrix} \quad (S36)$$

$$\gamma^2 = \begin{pmatrix} 0 & 0 & 0 & -i \\ 0 & 0 & i & 0 \\ 0 & i & 0 & 0 \\ -i & 0 & 0 & 0 \end{pmatrix} \quad (S37)$$

$$\gamma^3 = \begin{pmatrix} 0 & 0 & 1 & 0 \\ 0 & 0 & 0 & -1 \\ -1 & 0 & 0 & 0 \\ 0 & 1 & 0 & 0 \end{pmatrix} \quad (S38)$$

$$\sigma^{uv} = \frac{i}{2}[\gamma^\mu, \gamma^v], (\mu < \gamma) \tag{S39}$$

$$\gamma^5 = i\gamma^0\gamma^1\gamma^2\gamma^3 = \begin{pmatrix} 0 & 0 & 1 & 0 \\ 0 & 0 & 0 & 1 \\ 1 & 0 & 0 & 0 \\ 0 & 1 & 0 & 0 \end{pmatrix} \tag{S40}$$

$$V_0 = V_1 = V_2 = 0 \tag{S41}$$

$$V_3 = \frac{\gamma}{2}(k_x - ik_y)^2 - \frac{\gamma+k}{4}(k_x + ik_y)^2 \tag{S42}$$

$$T_{01} = -2i(\eta k^2 + \tau) \tag{S43}$$

$$T_{02} = \vartheta \tag{S44}$$

$$T_{03} = \frac{i}{2}(k - \gamma)(k_x + ik_y^2)^2 \tag{S45}$$

$$T_{12} = 2\theta_I(k) \tag{S46}$$

$$T_{13} = -2v_I k_y \tag{S47}$$

$$T_{23} = 2\alpha(k_x^2 - k_y^2) \tag{S48}$$

$$P = \frac{\gamma}{2}(k_x - ik_y)^2 + \frac{\gamma+k}{4}(k_x + ik_y)^2 \tag{S49}$$

$$A_0 = \frac{\gamma-k}{4}(k_x + ik_y)^2 \tag{S50}$$

$$A_1 = \gamma_I k_x \tag{S51}$$

$$A_2 = -2\alpha k_x k_y \tag{S52}$$

$$A_3 = 0 \tag{S53}$$

## II. The parameter space of the ext-kagome-II

Since shrinking and expanding uniformly toward the unit-cell center cannot naturally generate the double degeneracy of the Dirac point in ext-kagome-II, we seek to obtain the analytical solution of the Hamiltonian. We derive the analytical solution for the Hamiltonian at the Γ point of ext-kagome-II. Ext-kagome-II necessarily exhibits three pairs of doubly degenerate energy bands, corresponding to the six topmost bands in Fig. 3(b) and equation S(54). To identify another pair of doubly degenerate bands in the vicinity of the Dirac point, the relevant univariate cubic equation (S55) must develop a multiple root. At the Γ point, the Hamiltonian reduces to a cubic secular equation in energy, and the presence of an accidental degeneracy corresponds to the vanishing of the discriminant of this cubic equation. Then, by applying the standard criterion for multiple roots, we obtain the condition summarized in Eqs. (S56)–(S61). When the parameters do not satisfy this condition, the double-degenerate Dirac cone disappears.

$$\text{eqn1}: x^3 + J^2 K_1 + (2J + K_1)x^2 - JK_2^2 - JK_3^2 - 2K_2 K_3 K_4 - K_1 K_4^2 + (J^2 + 2JK_1 - K_2^2 - K_3^2 - K_4^2)x = 0 \tag{S54}$$

$$\text{eqn2}: x^3 + (-4J - 2K_1)x^2 - 8J^2 K_1 + 2JK_2^2 + 2JK_3^2 - 2K_2 K_3 K_4 + 2K_1 K_4^2 + (4J^2 + 8JK_1 - K_2^2 - K_3^2 - K_4^2)x = 0 \tag{S55}$$

$$a = 1 \tag{S56}$$

$$b = (-4J - 2K_1) \tag{S57}$$

$$c = (4J^2 + 8JK_1 - K_2^2 - K_3^2 - K_4^2) \tag{S58}$$

$$d = -8J^2 K_1 + 2JK_2^2 + 2JK_3^2 - 2K_2 K_3 K_4 + 2K_1 K_4^2 \tag{S59}$$

$$\delta_1 = -27a^2d^2 + 18abcd - 4b^3d - 4ac^3 + b^2c^2 \tag{S60}$$
$$\delta_2 = b^2 - 3ac \tag{S61}$$

When $\delta_1 = 0$, and $\delta_2 > 0$, doubly degeneracy appears at the Dirac point. As a consequence of these conditions, $K_2 = K_3$ emerges. However, a critical condition must be maintained: $K_2(K_3) \neq K_4$, to prevent the structure from reverting to the kagome structure.

### III. Mode profiles evolution with band inversion

The four eigenstates of the $H_\pm^{1st}$ exhibit a pairwise degeneracy at the Γ point $k = 0$. When applying the perturbation to the ext-kagome-I, the energy band structure opened. As the ratio of intracell coupling strength and intercell coupling strength varies, the band structure changes from opening through closing to reopening. The energy band, dipolar modes and quadrupolar modes were calculated by the COMSOL Multiphysics for both topologically nontrivial and trivial phases. The dipolar and quadrupolar modes reverse as band structure inversion occurs. The terms "dipolar modes" and "quadrupolar modes" refer to the spatial symmetry characteristics of the eigenmodes at the Γ point.

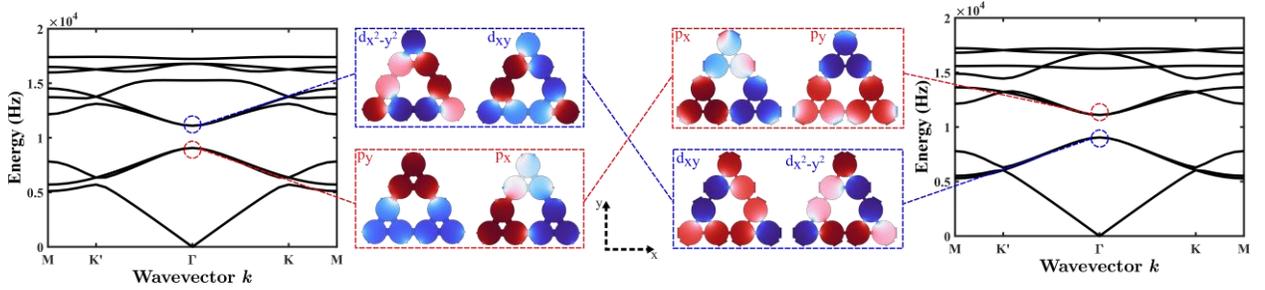

FIG. S5. Simulated energy band structure and mode profiles showing the sound pressure mode of ext-kagome-I unit cell for dipolar and quadrupolar eigenmodes for transition. The left figure is the trivial phase, and the right is the nontrivial phase.

### IV. The Wannier bands of the extended kagome

We calculate the Wannier center of the unit cell with different K(J) parameters by the TBM. Because of the band structure degeneracy in the system, we need to calculate all the bands with degenerations occurring below the F vicinity of the Dirac point. Meanwhile, we pay attention to the gauge choice for the maximal localized Wannier center. We use the Wilson loop along the base vector $k_1$ and $k_2$ to calculate the Wannier center. The calculation of the Wilson loop is as follows [5]:

$$W_{2\pi+k_s \leftarrow k_s, k_t} = \langle u_{2\pi+k_s,k_t}|u_{2\pi+k_s-\delta k,k_t}\rangle\langle u_{2\pi+k_s-\delta k,k_t}|u_{2\pi+k_s-2\delta k,k_t}\rangle \cdots \langle u_{k_s+2\delta k,k_t}|u_{k_s+\delta k,k_t}\rangle\langle u_{k_s+\delta k,k_t}|u_{k_s,k_t}\rangle \tag{S55}$$

where $k_s, k_t = 0, \delta k, \ldots, (N_k - 1)\delta k$, $\delta k = \frac{1}{N_k} \times \frac{4\pi}{\sqrt{3}a}$. According to formula S55, we calculate the Wannier bands of the two types of extended kagome. Because of the degenerate nature of the energy bands, we need a gauge choice for maximal localized Wannier center and simultaneously calculate Wannier bands for all degenerated energy bands below the vicinity of the Dirac point, as shown in Fig. S6.

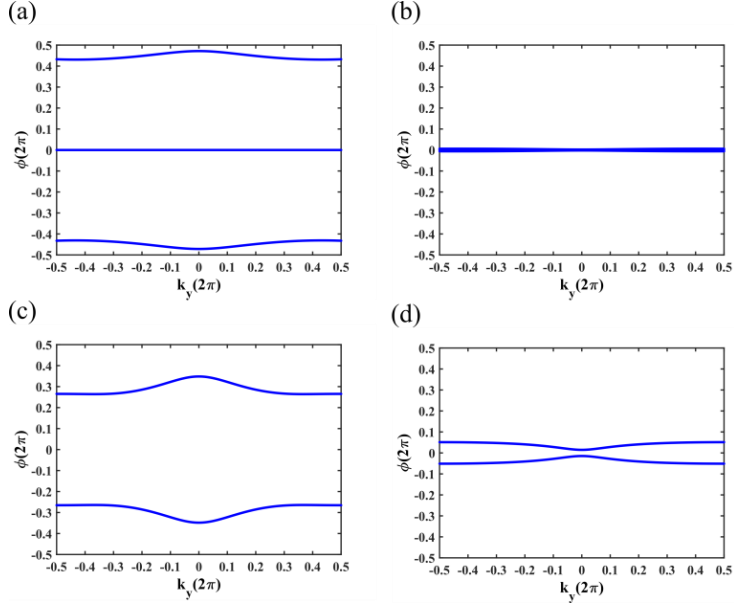

FIG. S6. Wanneir center of two types of extended kagome. (a) The topologically nontrivial ext-kagome-I. (b) The trivial ext-kagome-I. (c) The topologically nontrivial ext-kagome-II. (d) the trivial ext-kagome-II.

To further check the bulk polarization and boundary charge of the systems, we calculate the Wannier center for the two types of the extended kagome for the same parameters as the corner state phase (Fig. 6 and Fig. 7 in the main text), which is shown as Fig. S6. The Wannier center of ext-kagome-I is shown in Figs. S6(a) and S6(b), whose shape looks like the conventional kagome. The Figs. S6(a) and S6(b) correspond to the nontrivial phase and the trivial phase, respectively. Fig. S6(c) and (d) show the topologically nontrivial phase and the trivial phase of ext-kagome-II, respectively. The Wannier centers in Fig. S6 present the distinct characteristics of the extended kagome lattices to identify the topologically nontrivial and trivial phases, consistent with the $Z_2$ invariant calculation.

## V. The edge state profiles near the different corner states

As demonstrated in Fig. 6 of the main text, (c) and (d) have a series of edge states, respectively. Mode profile of edge state in Fig. S7(a) corresponding to the edge state in Fig. 6(a). Mode profile of edge state in Fig. S7(b) corresponding to the edge state in Fig. 6(b). They all clearly illustrate characteristics of edge states in this system.

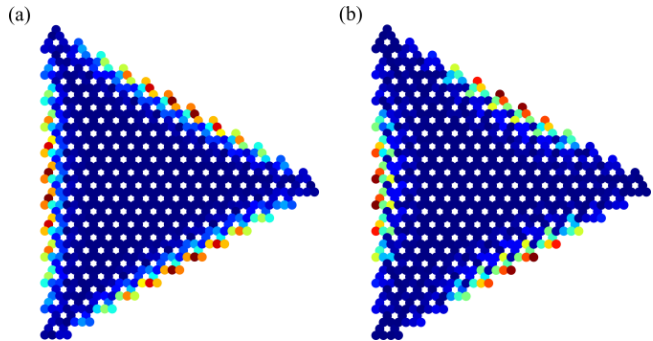

FIG. S7. Mode profile of edge states under different frequencies. (a) Mode profile of edge states in FIG. 6(c). (b)

Mode profile of edge states in FIG. 6(d)

## VI. The random perturbation of corner states

We first applied a random perturbation within [−15%, +15%] to the ext-kagome-I model. In this regime, all type-I corner states persist, as shown in panels (d)–(g) of Fig. S8. Notably, the lowest-frequency type-I corner states exhibit substantially stronger robustness compared with the remaining type-I corner states. Our tests further indicate that type-II corner states are less robust than type I, and can only be maintained under a much smaller perturbation window of [−1%, +1%] (see Fig. S9).

We then applied a random [−5%, +5%] perturbation to the ext-kagome-II model. All corner states remain present, as shown in Fig. S10. Consistent with the first model, lower-frequency corner states display stronger robustness.

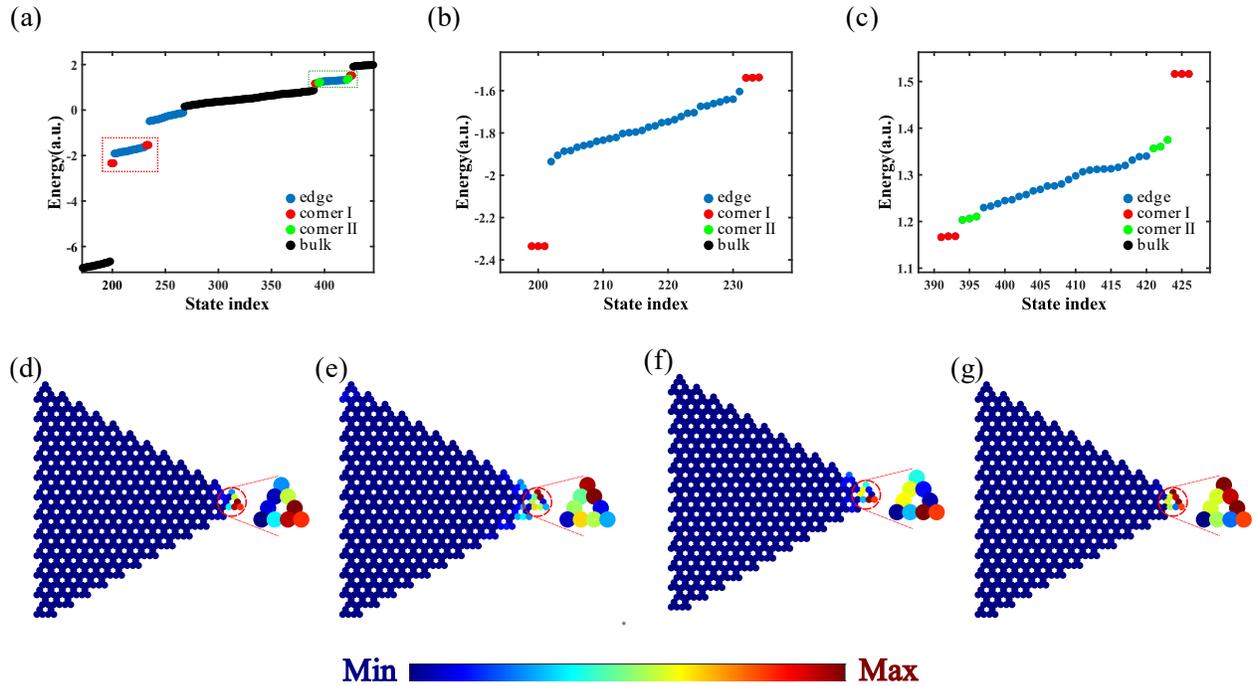

FIG. S8 Finite triangular ext-kagome-I structure after applying a random perturbation within [−15%, +15%]. (a) The energy spectrum of ext-kagome-I with a triangular-shaped array for $J/K = 5$. Black, blue, red and green dots represent the bulk states, edge states, type-I corner states and type-II corner states, respectively. (b) The enlarged image in the red dashed box in (a). (c) The enlarged image in the green dashed box in (a). (d-g) Mode profiles of type-I corner states with frequencies from low to high corresponding to the (a). The color bar shows the amplitude of the mode.

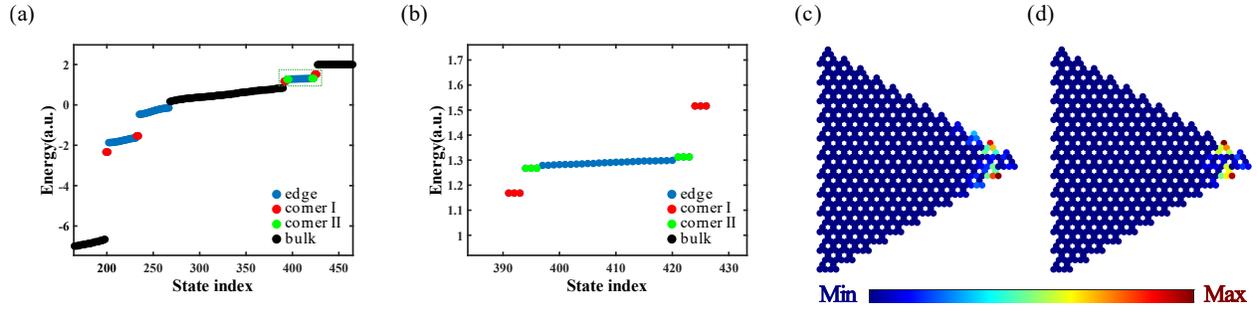

FIG. S9 Finite triangular ext-kagome-I structure after applying a random perturbation within [−1%, +1%]. (a) The energy spectrum of ext-kagome-I with a triangular-shaped array for $J/K = 5$. Black, blue, red and green dots represent the bulk states, edge states, type-I corner states and type-II corner states, respectively. (b) The enlarged image in the green dashed box in (a). (c-d) Mode profiles of type-II corner states with frequencies from low to high corresponding to the (a). The color bar shows the amplitude of the mode.

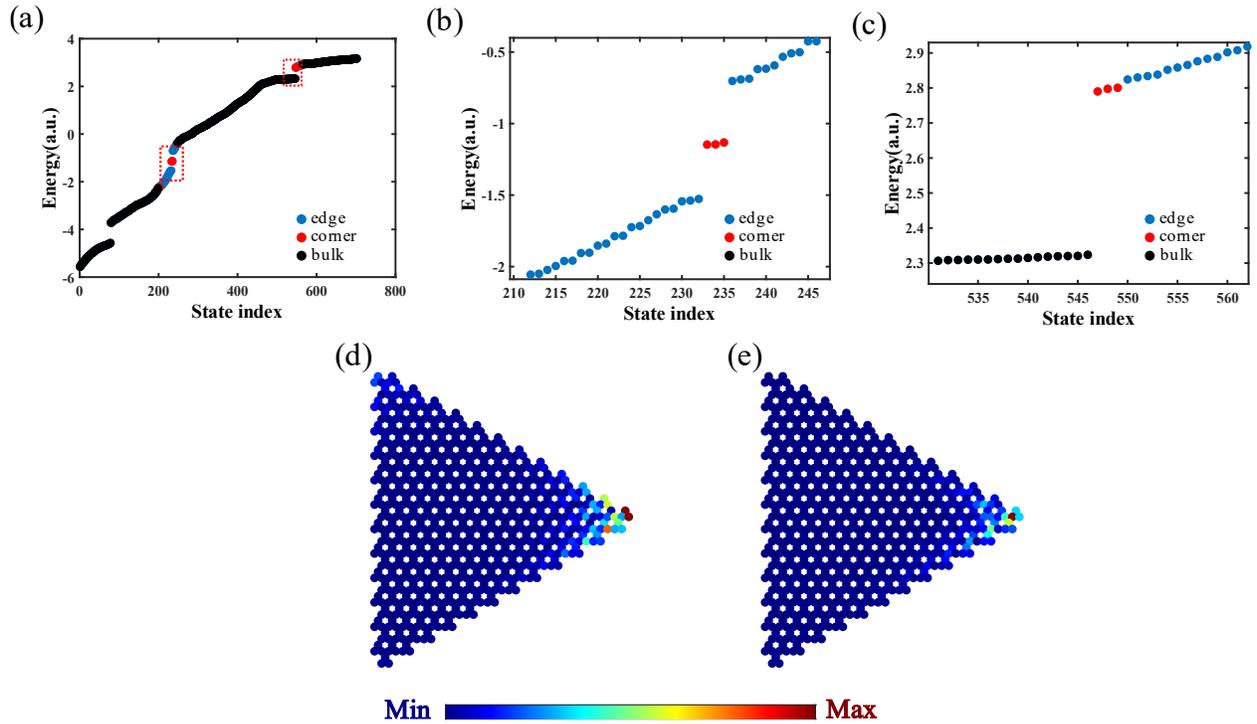

FIG. S10 Finite triangular ext-kagome-II structure after applying a random perturbation within [−5%, +5%]. (a) The energy spectrum of ext-kagome-II with a triangular-shaped array for $K_1 = -2.0$, $K_2 = -1.2$, $K_3 = -1.2$, $K_4 = -0.8$, $J = -1.5$. Black, blue, red and dots represent the bulk states, edge states and corner states, respectively. (b) The enlarged image in the lower red dashed box in (a). (c) The enlarged image in the higher red dashed box in (a). (d) Mode profiles of the corner states corresponding to the lower frequency in (b). (e) Mode profiles of the corner state corresponding to a higher frequency in (c). The color bar shows the amplitude of the mode profile in (d-e).